\documentclass[twocolumn,aps,superscriptaddress]{revtex4}

\usepackage{graphicx}
\usepackage{dcolumn}
\usepackage{bm}
\usepackage{amsmath,amssymb,amsfonts}
\usepackage{color}
\usepackage{braket}
\usepackage{mathtools}

\usepackage{siunitx}

\begin{document}

\title{Quantum Transport and Magnetism of Dirac Electrons in Solids}

\author{Hiroki Isobe}
\affiliation{Department of Applied Physics, University of Tokyo, Bunkyo, Tokyo, 113-8656 Japan}

\author{Naoto Nagaosa}
\affiliation{RIKEN Center for Emergent Matter Science (CEMS), Wako, Saitama 351-0198, Japan}
\affiliation{Department of Applied Physics, University of Tokyo, Bunkyo, Tokyo, 113-8656 Japan}

\begin{abstract}
The relativistic Dirac equation covers the fundamentals of electronic phenomena in solids and as such it effectively describes the electronic states of the topological insulators like Bi$_2$Se$_3$ and Bi$_2$Te$_3$.  Topological insulators feature gapless surface states and, moreover, magnetic doping and resultant ferromagnetic ordering break time-reversal symmetry to realize quantum anomalous Hall and Chern insulators.  Here we focus on the bulk and investigate the mutual coupling of electronic and magnetic properties of Dirac electrons.  Without carrier doping, spiral magnetic orders cause a ferroelectric polarization through the spin-orbit coupling.  In a doped metallic state, the anisotropic magnetoresistance arises without uniform magnetization.  We find that electric current induces uniform magnetization and conversely an oscillating magnetic order induces electric current.  Our model provides a coherent and unified description of all those phenomena. 
The mutual control of electric and magnetic properties demonstrates implementations of antiferromagnetic spintronics.   
We also discuss the stoichiometric magnetic topological insulator MnBi$_2$Te$_4$.
\end{abstract}

\maketitle

Relativistic effects on electrons, exemplified by the spin-orbit coupling, mingle the spin and orbital degrees of freedom and bring the interplay between electric and magnetic properties.  
Multiferroics is a manifestation in insulators, where a magnetization induces an electric polarization and vice versa \cite{multiferro1,multiferro2,multiferro3,multiferro4}.  
In metals and semiconductors, the spin-orbit coupling enables control of electrons' spin from electric current, and it is essential for spintronics.  For example, the Rashba spin-orbit coupling causes the Edelstein effect \cite{Edelstein}, which produces spin polarization by electric current in inversion-breaking systems.  Spintronics conventionally utilizes ferromagnets.  Antiferromagnetic spintronics recently has gained more interest owing to various advantages such as fast response, no stray field, and large magnetotransport effects \cite{AF_spintronics_1,AF_spintronics_2,AF_spintronics_3}.  However, because the net magnetization vanishes in an antiferromagnet, manipulation and detection remain essential challenges.

A magnetization pattern in general configures a spiral order with strong correlation or with magnetic elements with a fixed magnetic moment.  
Magnetism breaks time-reversal symmetry $\cal{T}$ even though a spiral magnetic order may have no net magnetic moment.  In addition, a spiral order is characterized by a wavevector $\bm{Q}$ and often breaks inversion symmetry $\cal{P}$ regardless of the underlying crystalline symmetry.

We study various phenomena related to broken $\cal{T}$ and $\cal{P}$ symmetries in magnetic Dirac materials in a unified fashion.  
The spin-orbit coupling naturally arises from the Dirac equation; as it abides by relativity, the coupling between the electric and magnetic degrees of freedom is contained.  
There are various materials where the Dirac Hamiltonian becomes the effective model near the chemical potential.  Examples are the three-dimensional topological insulators (TIs) Bi$_2$Se$_3$ and Bi$_2$Te$_3$ \cite{Xia,Zhang,Moore,Chen,Hsieh,Hor}.  With an insulating bulk, the topologically protected surface states determine the physical properties, which have been extensively studied \cite{TI1,TI2}.  In the doped case, however, the bulk states dominate the electric and magnetic properties of the system.  
When a magnetic order is present, the bulk of TIs offer an ideal laboratory to study the Dirac electrons with the exchange coupling to the magnetic moments.

In this work, we consider the electromagnetic response of a gapped Dirac system coupled to local magnetic moments.  
Our model describes the magnetically doped TIs Bi$_2$Se$_3$ and Bi$_2$Te$_3$ \cite{Cava,ZXShen,Hasan,Okada,Wang}, where the magnetic dopants couple locally to the Dirac electrons via the exchange coupling.  
We first show that an inversion-breaking magnetic order can generate a finite electric polarization in the insulating state while the pristine electronic system is centrosymmetric.  
In a doped metallic state, we reveal that a magnetic order can induce anisotropic resistance.  In addition, an electric field produces a uniform magnetization and in reverse an oscillating magnetic order generates direct current. 
We also discuss the intrinsic magnetic TI MnBi$_2$Te$_4$ \cite{Chulkov,Chulkov2,McQueeney,Xu} and the possibility of inversion-breaking magnetic orders in magnetic TIs.

\begin{figure*}
\centering
\includegraphics[width=0.9\hsize]{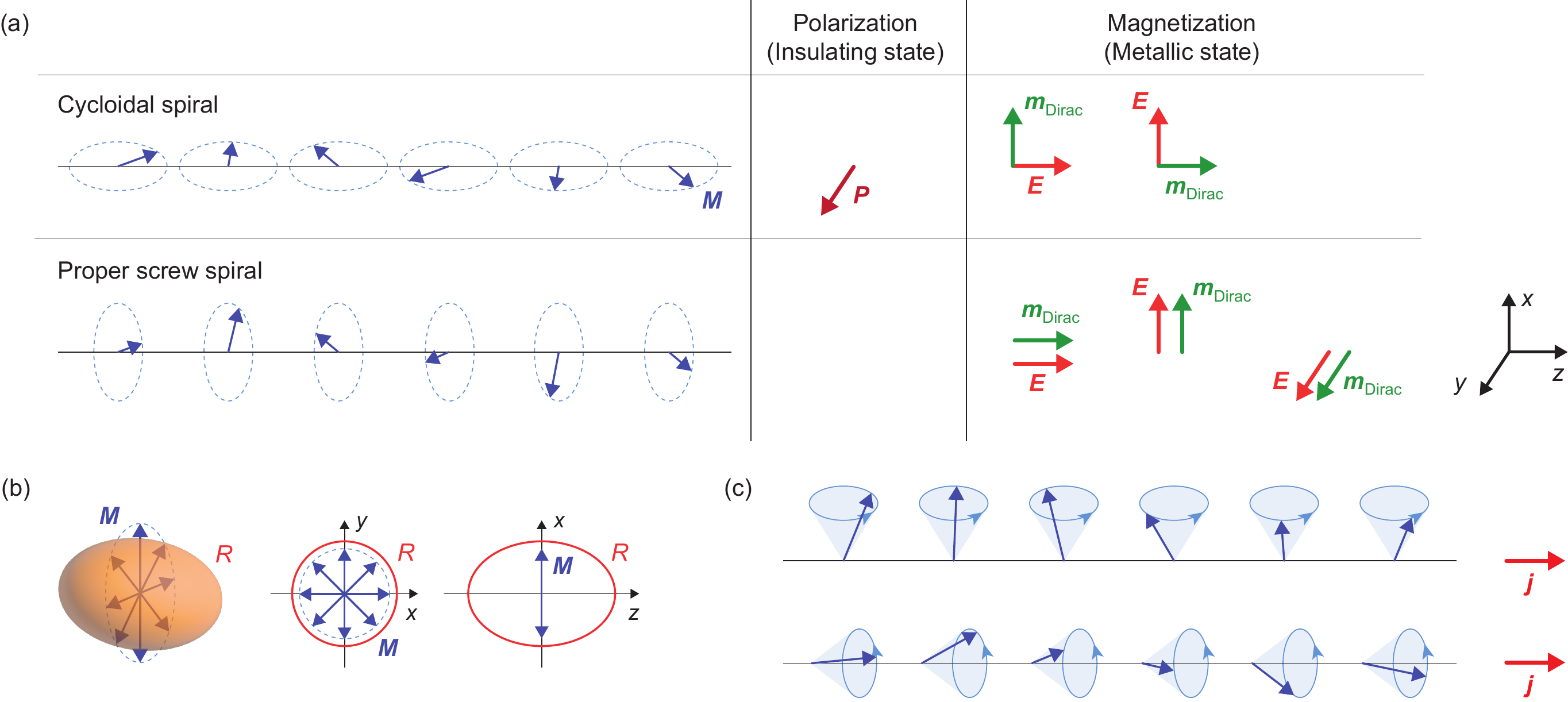}
\caption{Electromagnetic properties of a magnetic TI. 
(a) Polarization in the insulating state and magnetization in the metallic state induced by cycloidal and proper screw spiral magnetic orders.  
With the wavevector $\bm{Q} \parallel \hat{z}$, the cycloidal spiral order is characterized by $\bm{M}_{\bm{Q}} \propto \hat{y} - i\hat{z}$ and the proper screw spiral order by $\bm{M}_{\bm{Q}} \propto \hat{x} - i\hat{y}$.  
For those two spiral orders, only the cycloidal order displays a finite polarization according to Eq.~\eqref{eq:polarization}.  In the metallic state, the induced magnetization of the Dirac electrons $\bm{m}_\text{Dirac}$ varies with the electric field; see Eq.~\eqref{eq:magnetization}.  
(b) Anisotropic magnetoresistance in the presence of a spiral magnetic order Eq.~\eqref{eq:AMR}.  
The left panel is a three-dimensional illustration of the anisotropic resistance $R$ , and the center and right panels are the two distinct plane cuts, displaying the anisotropy in the plane perpendicular to the local magnetic moments.  
(c) Uniform direct current induced by oscillating magnetic orders.  For oscillations forming cycloidal and proper screw patterns, the generated direct current is parallel to the wavevector of the magnetic orders, following Eq.~\eqref{eq:current_magnetic}.  
}
\label{fig:results}
\end{figure*}

\textit{Model}: 
We consider a three-dimensional isotropic gapped Dirac system.  
Such an electronic system is realized, for example, in the bulk of TIs.  For the TIs Bi$_2$Se$_3$ and Bi$_2$Te$_3$, the energy bands near the $\Gamma$ point describe the low-energy behavior, which consists of the spin $\bm{\sigma}$ and $p$ orbitals $\bm{\tau}$ from Bi $(\tau_z = +1)$ and Se/Te $(\tau_z = -1)$.  To linear order in momentum $\bm{k}$, the $\bm{k}\cdot\bm{p}$ Hamiltonian becomes 
\begin{equation}
\label{eq:Dirac}
H_0(\bm{k}) = m\beta + \bm{\alpha}\cdot\bm{k}, 
\end{equation}
where the $4 \times 4$ matrices $\bm{\alpha} = \bm{\sigma}\tau_x$ and $\beta = \tau_z$ satisfy the anticommutation relations $\{ \alpha_a, \alpha_b \} = \{ \alpha_a, \beta \} = 0$ $(a \neq b)$ and $\alpha_a^2 = \beta^2 = I$ ($I$: identity matrix) \cite{Zhang}.  
We set $\hbar = 1$.  
The pristine system preserves inversion $\mathcal{P} = \tau_z$ and time reversal $\mathcal{T} = i\sigma_y\mathcal{K}$ with the complex conjugate operator $\mathcal{K}$: $\mathcal{P} H_0(\bm{k}) \mathcal{P}^{-1} = H_0(-\bm{k})$ and $\mathcal{T} H_0(\bm{k}) \mathcal{T}^{-1} = H_0(-\bm{k})$.  
The kinetic term renders the spin and orbital coupling, so that neither is a good quantum number. 
The sign of the mass can be either positive or negative, which describes the band inversion near the $\Gamma$ point.

Magnetic dopants such as Mn, Cr, and Fe can substitute the Bi sites of Bi$_2$Se$_3$ and Bi$_2$Te$_3$ \cite{magTI3}.  Their local magnetic moments break time-reversal symmetry and tend to form a magnetic order.  In a metallic state, the Ruderman--Kittel--Kasuya--Yosida (RKKY) interaction favors ferromagnetism when the Fermi level is near the Dirac point, but in general a complex magnetic order may occur depending on the Fermi level, anisotropy, and inhomogeneity \cite{order1,order2,Overhauser1,Overhauser2}.  An effective spin Hamiltonian reflecting such details of the system determines the magnetic order $\bm{M}(\bm{r}) = \sum_{\bm{Q}} \bm{M}_{\bm{Q}} e^{i\bm{Q}\cdot\bm{r}}$, which we take as given in the following analyses.

The exchange coupling yields the local magnetic coupling to the Dirac electrons.  We note that the exchange coupling is orbital dependent \cite{Wakatsuki}: 
\begin{align}
H'(\bm{r}) &= -J\bm{M}(\bm{r})\cdot\bm{\sigma} -J'\tilde{\beta}\bm{M}(\bm{r})\cdot\bm{\sigma}.   
\label{eq:exchange}
\end{align}
Here, we introduce $\tilde{\beta} = \tau_z \operatorname{sgn}(m)$ for later convenience.  
The two coupling constants $J$ and $J'$ describe the different strengths of the exchange coupling for the two orbitals $(\tau_z = \pm 1)$.

\textit{Insulating state}: 
The bulk is insulating when the chemical potential lies inside the mass gap.  While the electronic system preserves inversion, the magnetic order may violate it, allowing a finite electric polarization.  The calculation of the polarization follows the method by King-Smith and Vanderbilt \cite{King-Smith}.  We find an inversion-breaking magnetic order produces a finite polarization of the Dirac electrons
\begin{equation}
\Delta\bm{P} = -\frac{eJJ'}{6\pi^2 |m|} \sum_{\bm{Q}} \operatorname{Im} [\bm{M}_{\bm{Q}}^* (\bm{Q}\cdot\bm{M}_{\bm{Q}})]; 
\label{eq:polarization}
\end{equation}
see Supplemental Material (SM) for details \cite{SM}. 

The result conforms to the analyses of a Ginzburg--Landau model \cite{Mostovoy} and a microscopic model \cite{Katsura}, where certain chiral magnetic orders induce a finite electric polarization.  
For cycloidal and proper screw orders, only the former induce a finite polarization perpendicular to the wavevector in the magnetization plane according to Eq.~\eqref{eq:polarization} [Fig.~\ref{fig:results}(a)]. 
The product $JJ'$ implies that the strengths of the exchange coupling should be different for the two orbitals for a finite polarization.  The orbital-dependent exchange coupling mixes the conduction and valence bands by the magnetic order to realize a finite polarization.

\textit{Effective Hamiltonian in the metallic state}: 
When the system is metallic, we expect various responses to an external electromagnetic field.  As charges in the vicinity of the Fermi surface are dominantly responsible to electromagnetic response, it is convenient to derive the effective Hamiltonian for the bands that cross the Fermi energy.   
We obtain the effective Hamiltonian by following the method by Foldy and Wouthuysen \cite{FW}, and Tani \cite{Tani}, which we can calculate as a perturbative series in the large mass limit $|m| \gg |\epsilon_F|$ ($\epsilon_F$: the Fermi energy measured from a band edge) \cite{SM}.  
In the presence of an external electromagnetic field, the effective Hamiltonian to order $m^{-2}$ is  
\begin{widetext}
\begin{align}
H_\text{eff} &= |m|\tilde{\beta} -e\Phi - (J + J'\tilde{\beta})\bm{M}\cdot\bm{\sigma} + \frac{\tilde{\beta}}{2|m|}(\bm{\Pi}\cdot\bm{\Pi}+e\bm{\sigma}\cdot\bm{B}) 
+ \frac{e}{8m^2}(\nabla\cdot\bm{E}) + \frac{e}{8m^2}[\bm{\Pi}\cdot(\bm{\sigma}\times\bm{E})+(\bm{\sigma}\times\bm{E})\cdot\bm{\Pi}] \nonumber\\
&\quad +\frac{J}{8m^2} \{ (\bm{\Pi}\cdot\bm{\sigma})[-i\nabla\cdot\bm{M} + \bm{\sigma}\cdot(\nabla\times\bm{M}) - 2i(\bm{\sigma}\times\bm{M})\cdot\bm{\Pi}] +\text{H.c.} \} \nonumber\\
&\quad +\frac{J'\tilde{\beta}}{8m^2} \{ (\bm{\Pi}\cdot\bm{\sigma})[-i\nabla\cdot\bm{M} + \bm{\sigma}\cdot(\nabla\times\bm{M}) +2\bm{M}\cdot\bm{\Pi}] + \text{H.c.} \} 
\label{eq:effectiveH}
\end{align}
\end{widetext}
with $\bm{\Pi} = \bm{p} + e\bm{A}$ and the momentum operator $\bm{p} = -i\nabla$.  The charge of an electrons is $-e$.  
The electric and magnetic fields are $\bm{E} = -\nabla \Phi - \partial\bm{A}/\partial t$ and $\bm{B} = \nabla\times\bm{A}$, respectively, with the scalar potential $\Phi$ and the vector potential $\bm{A}$.  
In the effective Hamiltonian, $\tilde{\beta} = \pm 1$ signifies the energy bands: $\tilde{\beta} = +1$ corresponds to the conduction band and $\tilde{\beta} = -1$ to the valence band.  
Although we originally define $\tilde{\beta} = \tau_z \operatorname{sgn}(m)$, it does not precisely label the orbitals after the unitary transformation.  

The last two terms of the effective Hamiltonian \eqref{eq:effectiveH} reveal the nontrivial coupling between the Dirac electrons and the magnetic order, which is central to the following results.  It manifests the strong spin-orbital coupling embedded in the Dirac Hamiltonian along with the exchange coupling.  
It also modifies the current density operator $\bm{\mathcal{J}} = ie [\bm{r}, H_\text{eff}]$ to become 
\begin{align}
&\ \bm{\mathcal{J}} \nonumber\\
=& -\frac{e}{m} \beta \bm{p} -\frac{e}{4m^2} (J+J'\beta)(\nabla\times\bm{M})
\nonumber\\
& - \frac{e}{4m^2} \Big\{ J\left[ -2i \bm{M}\times\bm{p}
- \bm{\sigma}\times(\bm{M}\times\bm{p})
- \bm{M}\times(\bm{\sigma}\times\bm{p}) \right] \nonumber\\
& \qquad +J'\tilde{\beta} \left[ \bm{\sigma}(\bm{M}\cdot\bm{p}) +\bm{M}(\bm{\sigma}\cdot\bm{p})\right] + \text{H.c.} \Big\} 
\label{eq:current_operator}
\end{align}
at zero frequency.  
The second term with $\nabla\times\bm{M}$ has a classical analog to the Amp\`ere's circuital law. 
The third term contains the local magnetic moment $\bm{M}$ and the spin of the Dirac electrons $\bm{\sigma}$.  It implies the possibility of the mutual control of the electric and magnetic degrees of freedom as we will see below.

\textit{Current under an electric field}: 
We perform perturbative calculations using functional derivatives to calculate response.  
We define the action $S = T\sum_{\omega_n} \int d\bm{r} \bar{\psi} (-i\omega_n + H_\text{eff}) \psi$, where $T$ is the temperature and $\omega_n = (2n+1)\pi T$ is the fermionic Matsubara frequency.  
Using the partition function $Z = \int D\bar{\psi} D\psi e^{-S}$, we obtain the current response in the presence of an external electric field $\bm{E}(\omega) = i\omega \bm{A}(\omega)$ $(\Phi=0)$ as 
\begin{equation}
j_a(\omega) = \langle \hat{j}_a(\omega) \rangle = \frac{1}{i\omega} \frac{\delta^2 \ln Z}{\delta A_a(-\omega) \delta A_b(\omega)} \Bigg|_{E=0} E_b(\omega).  
\label{eq:current}
\end{equation}
We note that it is equivalent to the Kubo formula.  
We calculate it perturbatively with respect to the exchange couplings $J$, $J'$, and the inverse mass $m^{-1}$, using the unperturbed Green's function $G_0 = (\omega - H_\text{eff}^0 -\Sigma)^{-1}$ with $H_\text{eff}^0 = m\beta + \beta k^2/(2m)$ and the self-energy $\Sigma$.  We approximate $\Sigma \approx -i\operatorname{sgn}(\omega_n)/(2\tau)$ with a constant $\tau$ to describe momentum relaxation in diffusive transport.  

The magnetic order alters the current flow.  When we focus on a spatially uniform current, the lowest-order corrections by the magnetic order appear as a product of $\bm{M}_{\bm{Q}}$ and $\bm{M}_{\bm{Q}}^*$.  By differentiating Eq.~\eqref{eq:current} with respect to $\bm{M}_{\bm{Q}}$ and $\bm{M}_{\bm{Q}}^*$, we obtain the conductivity tensor 
\begin{align}
\sigma_{ab}(\omega) 
&= \sigma_0(\omega) \delta_{ab} 
+ \sigma^\text{AH}_{ab}(\omega)
- \eta(\omega) \sum_{\bm{Q}} |\bm{M}_{\bm{Q}}|^2 \delta_{ab} \nonumber\\
&\quad + \eta'(\omega) \sum_{\bm{Q}} \left( M^*_{\bm{Q},a} M_{\bm{Q},b} + M_{\bm{Q},a} M^*_{\bm{Q},b} \right),
\label{eq:AMR}
\end{align}
where the coefficients are given by 
\begin{gather*} 
\sigma_0(\omega) = \frac{e^2 |n(\epsilon_F)| \tau_\omega}{|m|}, \\
\eta(\omega) = \frac{2e^2 |n(\epsilon_F)| \tau_\omega^3}{|m|} (J+J'\tilde{\beta})^2, \\
\eta'(\omega) = \frac{e^2 |n(\epsilon_F)| \tau_\omega}{8|m|^3} (-3J^2 + 5J'^2 - 2JJ'\tilde{\beta}).  
\end{gather*}
$n(\epsilon_F) \propto |\epsilon_F|^{3/2}$ is the carrier density ($n>0$ for electrons and $n<0$ for holes).  We introduce $\tau_\omega = \tau/(1-i\omega\tau)$ and retain the leading-order contributions in $m^{-1}$ in the expressions of $\sigma$, $\eta$, and $\eta'$.  
When there is a uniform magnetization $\bm{M}_{\bm{0}} \neq \bm{0}$, it yields the anomalous Hall contribution $\sigma^\text{AH}_{ab} \propto \varepsilon_{abc} M_{\bm{0},c}$; see SM for details \cite{SM}.  
$\bm{j}(\omega)$ depends only on $\bm{M}_{\bm{Q}}$ but does not directly depend on $\bm{Q}$ to this order.

The spatial pattern of the magnetic order modifies the conductivity at second order in $\bm{M}$.   
The first correction with $\eta(\omega)$ reduces the longitudinal conductivity, arising from the exchange coupling $- (J + J'\tilde{\beta})\bm{M}\cdot\bm{\sigma}$.  
The effect is isotropic and it does not require the spin-orbital coupling inherent in the Dirac Hamiltonian.  It resembles the magnetoresistance whereas there is no uniform magnetization by assumption.  
On the other hand, the $\eta'$ term can be traced to the coupling between the magnetic order and current, as we have seen in Eq.~\eqref{eq:current_operator}.  It gives rise to anisotropic corrections to the conductivity tensor $\sigma_{ab}(\omega)$ depending on the magnetic order.  

The second-order corrections to the conductivity correspond to the anisotropic magnetoresistance and the planar Hall effect \cite{AMR}.  
Both cycloidal and screw magnetic orders show the anisotropic resistance $R$ [Fig.~\ref{fig:results}(b)]: 
when the magnetic order lies in the $xy$ plane, the resistance is different in the $xy$ plane and along the $z$ axis.  
We emphasize, however, that the second-order effect in Eq.~\eqref{eq:AMR} appears even without a uniform magnetization.  Therefore, when there is no uniform magnetization, namely, $\sigma^\text{AH}_{ab}=0$, the conductivity tensor is symmetric: $\sigma_{ab}(\omega) = \sigma_{ba}(\omega)$.  
On the other hand, the anomalous Hall contribution is antisymmetric: $\sigma^\text{AH}_{ab}(\omega) = -\sigma^\text{AH}_{ba}(\omega)$ \cite{AHE}.

\textit{Magnetization by an electric field}: 
From the coupling between the current and the spin degrees of freedom, we expect that an electric field produces a finite magnetization of Dirac electrons even when the magnetic order has no uniform magnetization.  We evaluate the spin expectation value of the Dirac electrons $\langle \bm{\sigma} \rangle$ in the presence of an external electric field $\bm{E}$ and the magnetic order $\bm{M}$ using $H_\text{eff}$.  The uniform magnetization of the Dirac electrons is given by $\bm{m}_\text{Dirac} = -g\mu_B \langle \bm{\sigma} \rangle / 2$, where $g$ is the $g$-factor and $\mu_B$ is the Bohr magneton.  A perturbative calculation finds a finite magnetization under a static and uniform external electric field \cite{SM}
\begin{align}
\bm{m}_\text{Dirac} &= 
\lambda^{(1)} \sum_{\bm{Q}} (\bm{Q}\cdot\bm{E}) \operatorname{Im}(\bm{M}_{\bm{Q}}\times\bm{M}^*_{\bm{Q}}) \nonumber\\
&\quad + \lambda^{(2)} \sum_{\bm{Q}} \Big\{\! \operatorname{Im}[(\bm{M}^*_{\bm{Q}}\times\bm{Q})(\bm{M}_{\bm{Q}} \cdot \bm{E})] \nonumber\\
&\hspace{50pt} + \operatorname{Im}[(\bm{M}^*_{\bm{Q}}\times\bm{E})(\bm{Q}\cdot\bm{M}_{\bm{Q}})] \Big\}
\label{eq:magnetization}
\end{align}
with 
\begin{gather*}
\lambda^{(1)} = \frac{g\mu_B e n(\epsilon_F)}{m^2} \tau^3 J(J+J'\tilde{\beta}), \\
\lambda^{(2)} = \frac{g\mu_B e n(\epsilon_F)}{2 m^2} \tau^3 (J^2-J'^2).
\end{gather*}
Since the magnetization and the electric field transforms differently under inversion, an inversion-breaking magnetic order is necessary to induce magnetization by an electric field [Fig.~\ref{fig:results}(a)].  It allows detection of an inversion-breaking magnetic order through the magnetization by applying an electric field.  The change of the magnetization under an electric field can be attributed to $\bm{m}_\text{Dirac}$.  
The effect resembles the Edelstein effect but it appears in the bulk of a TI, where inversion is broken by a magnetic order.    
The extension to a time-dependence case is straightforward \cite{SM}.

\textit{Current by an oscillating magnetic order}: We now investigate whether an external magnetic field induces an electric current.  The magnetic field should vary in time as a spatially uniform current cannot exist in the equilibrium.  The external magnetic field applied to a metallic system with a magnetic order has the following two effects: it couples to the itinerant electrons to induce cyclotron motion; at the same time, it drives the Rabi oscillation and the Larmor precession of the local magnetic moments.  

We first check if a uniform oscillating magnetic field $\bm{B}(\omega)$ induces a uniform current in the presence of a static magnetic order.  
From the symmetry consideration, the lowest-order contribution should have the form $j_a(\omega) = \kappa_{abcd} B_b(\omega) M_{\bm{Q},c} M_{-\bm{Q},d}$ with $\kappa_{abcd}$ linear in $Q$.  
However, this mechanism is improbable.  The conductivity tensor Eq.~\eqref{eq:AMR} is insensitive to inversion breaking, so that the cyclotron motion of the Dirac electrons would not yield a uniform current.  We calculate $\kappa_{ijkl}$ perturbatively and observe that it vanishes to order $Q J^2 n(\epsilon_F)/m^2$ \cite{SM}.

We then examine current response by an oscillating magnetic order.  
If it is finite, an external magnetic field induces an electric current by making the local magnetic moments oscillate.  
We write the spatial and temporal dependence of the magnetic order as $\bm{M}(\bm{r},t) = \sum_{\bm{Q}\omega} \bm{M}_{\bm{Q}\omega} e^{i(\bm{Q}\cdot\bm{r}-\omega t)}$.  Here, we seek the uniform current response of the form $j_a(\omega_1+\omega_2) = \gamma_{abc}(\omega_1,\omega_2,\bm{Q}) M_{\bm{Q}\omega_1,b} M_{-\bm{Q}\omega_2,c}$, where $\gamma_{abc}$ is linear in the wavevector $\bm{\bm{Q}}$ to capture the inversion breaking by the magnetic order and hence to comply with the symmetry constraint. 
As a second-order response, the output frequency is the sum of two input frequencies.  
We can calculate the current response similarly to $\sigma_{ab}(\omega)$ \cite{SM}: 
\begin{align}
\bm{j}(\omega) 
= &\sum_{\bm{Q}\omega_1\omega_2} \delta_{\omega_1+\omega_2,\omega} \Big[ \gamma^{(S)}(\omega_1,\omega_2) \bm{Q}\times(\bm{M}_{1}\times\bm{M}_{2}) \nonumber\\
&\quad + \gamma^{(A)}(\omega_1,\omega_2) \nonumber\\
&\hspace{15pt} \times \{ J [\bm{M}_{1}\times(\bm{Q}\times\bm{M}_{2}) + \bm{M}_{2}\times(\bm{Q}\times\bm{M}_{1})] \nonumber\\
&\hspace{20pt} + J'\tilde{\beta} [\bm{M}_{1}(\bm{Q}\cdot\bm{M}_{2}) + \bm{M}_{2}(\bm{Q}\cdot\bm{M}_{1})] \} \Big],
\label{eq:current_magnetic}
\end{align}
where we denote $\bm{M}_1 = \bm{M}_{\bm{Q}\omega_1}$, $\bm{M}_2 = \bm{M}_{-\bm{Q}\omega_2}$, and the coefficients are 
\begin{gather*}
\begin{aligned}
\gamma^{(S)}(\omega_1,\omega_2) &= \frac{e}{8m^2} n(\epsilon_F) (J+J'\tilde{\beta})^2 \\
&\quad \times i(\omega_1+\omega_2)\tau_{\omega_1+\omega_2} ( \omega_1 \tau_{\omega_1}^2 + \omega_2 \tau_{\omega_2}^2 ),
\end{aligned}
\\
\gamma^{(A)}(\omega_1,\omega_2) = -\frac{e}{4m^2} (J+J'\tilde{\beta}) n(\epsilon_F) ( \omega_1 \tau_{\omega_1}^2 - \omega_2 \tau_{\omega_2}^2 ). 
\end{gather*}
$\gamma^{(S)}(\omega_1,\omega_2)$ and $\gamma^{(A)}(\omega_1,\omega_2)$ are symmetric and antisymmetric under the exchange of $\omega_1$ and $\omega_2$, respectively.  
$\gamma^{(A)}(\omega,-\omega)$ corresponds to zero-frequency response, namely direct current depicted in Fig.~\ref{fig:results}(c), and $\gamma^{(S)}(\omega,\omega)$ to $2\omega$ response.   

It is worth contrasting the current response in the metallic state \eqref{eq:current_magnetic} with the polarization in the insulating state \eqref{eq:polarization} as they reflect different material properties.  First, the current response requires dynamics of the magnetic order whereas the polarization is a thermodynamic quantity defined in the equilibrium.  The diffusive nature of the current is manifested in the appearance of the lifetime $\tau$.  Second, the current is carried by electric charges near the Fermi energy and it is thus proportional to the carrier density.  On the other hand, the polarization only involves the quantities that characterize the system, implying that it requires the information of the entire band structure.  Indeed, we cannot obtain Eq.~\eqref{eq:polarization} from the effective Hamiltonian \eqref{eq:effectiveH} but from the original model \eqref{eq:Dirac}.

As we have discussed, the local magnetic moments oscillate under a time-dependent external magnetic field to induce a uniform electric current. 
When the oscillation is near resonance, we may expect a larger current response.  
Since it is a second-order response with respect to the magnetic order, the response should be peaked at the zero frequency and double the resonance frequency.  
A magnetic order might also be driven by the spin wave spectroscopy technique \cite{spin-wave-1,spin-wave-2,Seki}.  
An oscillating magnetic field is induced by periodically aligned wave guides whereby the wavevector of the magnetic field is designed.  

\textit{Discussions}: 
We have revealed that electromagnetic response of a magnetic TI manifests the entanglement of the spin and orbital degrees of freedom and hence the electric and magnetic properties.  Particularly with an inversion-breaking magnetic order, it allows a measurement of electric properties through a magnetic probe and vice versa, and suggest applications in spintronics.  

In addition to the magnetically doped TIs, we can consider the stoichiometric magnetic TI MnBi$_2$Te$_4$.  It consists of stacking layers of TI films, bound by the van der Waals interaction \cite{Chulkov}.  The low-energy effective Hamiltonian is $H_\text{STI}(\bm{k}) = m \tau_x + v \tau_z (\hat{z}\times\bm{\sigma})\cdot\bm{k}_\perp + v_z k_z \tau_y$, where the stacking direction is set along the $z$ direction and $\tau_z$ corresponds to the top and bottom TI surface states of a constituent layer \cite{MacDonald}.  
In the SM \cite{SM}, we confirm that an inversion-breaking magnetic order induces an electric polarization in the insulating state, and derive the effective Hamiltonian for the metallic case to see that the current operator is affected by the magnetic order.

Experimentally, a spiral magnetic order has not yet been reported in magnetic TIs, but yet some experiments reveal noncollinear magnetic orders.   
Stacking layers of MnBi$_2$Te$_4$ realize a canted antiferromagnetic order \cite{Ovchinnikov}, and alternating stacks of MnBi$_2$Te$_4$ and Bi$_2$Te$_3$ lead to a variety of heterostructures (MnBi$_2$Te$_4$)$_m$(Bi$_2$Te$_3$)$_n$ \cite{Hesjedal}.  
The topological Hall effect is observed in the magnetic/non-magnetic topological insulator heterostructures Cr$_x$(Bi$_{1-y}$Sb$_y$)$_{2-x}$Te$_3$/(Bi$_{1-y}$Sb$_y$)$_2$Te$_3$ and a theory attributed its origin to a N\'eel-type skyrmion, consisting of the superposition of the local three spiral orders \cite{Yasuda}.  The topological Hall effect attributed to skyrmions is also observed in Mn-doped Bi$_2$Te$_3$ topological insulator films \cite{Wang2}.  Those observations suggest that various magnetic orders may appear by different stacks and material compositions.

We now estimate the magnitude of the effects that we have discussed using the material parameters of Cr$_x$(Bi$_{1-y}$Sb$_y$)$_{2-x}$Te$_3$ \cite{Yasuda}: $m = \SI{-300}{meV}$, $J = \SI{-5}{meV}$, $J' = \SI{1}{meV}$, and the velocity $v = \SI{5.0e5}{m/s}$; see SM for details \cite{SM}.  The magnetic moment per Cr atom is $M \approx 3\mu_B$.  We set $\epsilon_F = -\SI{100}{meV}$.  The RKKY interaction would form a magnetic order in the metallic state with the wavenumber $Q = 2k_F \approx \SI{1.5e9}{m^{-1}}$.  We estimate $\tau \approx \SI{5e-15}{s}$ from the longitudinal conductivity $\SI{100}{\ohm^{-1}.\cm^{-1}}$ with $|n| \approx \SI{1.4e19}{cm^{-3}}$.  
Then, the corrections to the conductivity are $-2\eta M^2 \approx \SI{-8.7}{\ohm^{-1}.\cm^{-1}}$ and $4\eta' M^2 \approx \SI{-0.4}{\ohm^{-1}.\cm^{-1}}$ for the isotropic and anisotropic parts, respectively.  
The magnetization induced by the current density $j = \SI{e8}{A/m}$ is $m_\text{Dirac} \sim 10^{-4}\,\mathrm{A/m}$.  
The current densities generated by an oscillating magnetic order at \SI{1}{GHz} are $\gamma^{(S)}QM^2 \approx \SI{9.2}{A/m^2}$ for the sum frequency generation and $\gamma^{(A)} JQM^2 \approx \SI{-2.3e5}{A/m^2}$.  We note that the former grows quadratically with frequency while the latter does linearly.  In the insulating state, the electric polarization is $\Delta P \approx \SI{1.8}{\micro\coulomb/m^2}$ with the same $Q$.  From those estimates, the electronic response is more likely to be observable that the magnetic one.

In addition to magnetic TIs, we also anticipate similar current response in magnetic Weyl and Dirac semimetals, where an emergent electromagnetic field plays a role as well as the Berry curvature \cite{Frohlich,Zaanen,Araki}.  
A surface, an interface, and a domain wall geometrically break inversion, and thus the existence of a magnetic order can also induce various response.  Such structures without inversion support the Dzyaloshinskii--Moriya interaction, which could contribute to a chiral magnetic order to reveal the effects that we have discussed.

\textit{Acknowledgment}: 
This work was supported by JST CREST Grant No. JPMJCR1874, Japan, and JSPS KAKENHI Grant No. 18H03676.

\onecolumngrid
\clearpage

\setcounter{section}{0}
\setcounter{equation}{0}
\setcounter{figure}{0}
\def\thesection{S\arabic{section}}
\def\theequation{S\arabic{equation}}
\def\thefigure{S\arabic{figure}}

\begin{center}
{\bf\large
	Supplemental Material
}
\end{center}

In Supplemental Material (SM), we describe the unitary transformation in the large mass limit, the details about the model for magnetically-doped topological insulators (TIs), the calculations of various response in the metallic state and polarization in the insulating state.  We also include the model and analysis of the model for the stoichiometric TI MnBi$_2$Te$_4$.  
We set $\hbar = 1$ unless otherwise noted.

\section{Effective Hamiltonian from a unitary transformation}

We derive the effective Hamiltonian for the Dirac system coupled to the magnetic order.  To this end, we perform a unitary (Foldy--Wouthuysen--Tani) transformation that diagonalizes the Hamiltonian in the orbital space in the large Dirac mass limit.  

We decompose the Hamiltonian as 
\begin{align}
H = m \beta + \mathcal{E} + \mathcal{O}, 
\end{align}
where $\mathcal{E}$ and $\mathcal{O}$ are the diagonal and off-diagonal terms in the orbital components:
\begin{gather}
\mathcal{E} = -e\Phi + H', \\
\mathcal{O} = \bm{\alpha}\cdot\bm{\Pi}. 
\end{gather}
$H'$ denotes the exchange coupling of the Dirac electrons to the magnetic order.  Here, we include the electromagnetic potential $(\Phi, \bm{A})$ and minimal coupling gives the canonical momentum $\bm{\Pi} = \bm{p} + e\bm{A}$.  We note that the diagonal part commutes with the matrix $\beta$ and that the off-diagonal part anticommutes with $\beta$:
\begin{equation}
[\beta,\mathcal{E}] = 0, \quad \{\beta,\mathcal{O}\} = 0, 
\end{equation}
which the following algebra relies on. 

An unitary transformation $e^{iS_1}$ with a Hermitian operator $S_1$ converts the Hamiltonian into 
\begin{align}
H_1 &= e^{iS_1} \left(H - i\frac{\partial}{\partial t} \right) e^{-iS_1} \nonumber\\
&= H + i[S_1,H] + \frac{i^2}{2!}[S_1,[S_1,H]] + \frac{i^3}{3!}[S_1,[S_1,[S_1,H]]] + \cdots \nonumber\\
&\quad + i(i\dot{S}_1) + \frac{i^2}{2!}[S_1,i\dot{S}_1] + \frac{i^3}{3!}[S_1,[S_1,i\dot{S}_1]] + \cdots \nonumber\\
&= m\beta + \mathcal{E}_1 + \mathcal{O}_1. 
\end{align}
$\mathcal{E}_1$ and $\mathcal{O}_1$ are the diagonal and off-diagonal terms in the orbital components after the unitary transformation. 
We determine $S_1$ to remove the off-diagonal terms at order $m^0$, requiring 
\begin{equation}
\mathcal{O} + i[S_1,m\beta] = 0.  
\end{equation}
The condition leads to 
\begin{equation}
S_1 = -\frac{i}{2m} \beta\mathcal{O}, 
\end{equation}
and thus the series expansion of the unitary transformation corresponds to the expansion with respect to the inverse of the Dirac mass $m^{-1}$.  
After performing the unitary transformation $e^{iS_1}$, we obtain 
\begin{align}
H_1 &= m\beta 
+ \mathcal{E} 
+ \frac{1}{2m}\beta\mathcal{O}^2 + \frac{1}{2m}\beta([\mathcal{O},\mathcal{E}]+i\dot{\mathcal{O}}) 
- \frac{1}{8m^2}[\mathcal{O},[\mathcal{O},\mathcal{E}]+i\dot{\mathcal{O}}] - \frac{1}{3m^2}\mathcal{O}^3 \nonumber\\
&\quad - \frac{1}{8m^3}\beta\mathcal{O}^4 - \frac{1}{48m^3}\beta[\mathcal{O},[\mathcal{O},[\mathcal{O},\mathcal{E}]+i\dot{\mathcal{O}}]]
+ O(m^{-4}), 
\end{align}
and the diagonal and off-diagonal terms in the orbital components $\mathcal{E}_1$, $\mathcal{O}_1$ are 
\begin{gather}
\mathcal{E}_1 = \mathcal{E} + \frac{1}{2m}\beta\mathcal{O}^2 - \frac{1}{8m^2}[\mathcal{O},[\mathcal{O},\mathcal{E}]+i\dot{\mathcal{O}}] - \frac{1}{8m^3}\beta\mathcal{O}^4 + O(m^{-4}), \\
\mathcal{O}_1 = \frac{1}{2m}\beta([\mathcal{O},\mathcal{E}]+i\dot{\mathcal{O}}) - \frac{1}{3m^2}\mathcal{O}^3 - \frac{1}{48m^3}\beta[\mathcal{O},[\mathcal{O},[\mathcal{O},\mathcal{E}]+i\dot{\mathcal{O}}]] + O(m^{-4}). 
\end{gather}
$\mathcal{E}_1$ and $\mathcal{O}_1$ again satisfy the same commutation and anticommutation relations 
\begin{equation}
[\beta,\mathcal{E}_1] = 0, \quad \{\beta,\mathcal{O}_1\} = 0. 
\end{equation}

We can iterate the same procedure to eliminate off-diagonal components at every order in $m$.  At the $j$-th repetition, the unitary transformation $e^{iS_j}$ leads to the Hamiltonian 
\begin{equation}
H_j = e^{iS_j} H_{j-1} e^{-iS_j} = m\beta + \mathcal{E}_j + \mathcal{O}_j, 
\end{equation}
where the Hermite operator $S_j$ is 
\begin{equation}
S_j = -\frac{i}{2m}\beta\mathcal{O}_{j-1}. 
\end{equation}
Here we consider $H_4$, which is diagonal to order $m^{-3}$:
\begin{gather}
\label{eq:S_unitary_4}
\mathcal{E}_4 = \mathcal{E} + \frac{1}{2m}\beta\mathcal{O}^2 - \frac{1}{8m^2}[\mathcal{O},[\mathcal{O},\mathcal{E}]+i\dot{\mathcal{O}}] - \frac{1}{8m^3}\beta\mathcal{O}^4 - \frac{1}{8m^3}\beta([\mathcal{O},[\mathcal{O},\mathcal{E}]+i\dot{\mathcal{O}}])^2 + O(m^{-4}), \\
\mathcal{O}_4 = O(m^{-4}).  
\end{gather}

To further calculate the expression, we define the matrix $\bm{\Sigma}$ as 
\begin{equation}
[ \alpha_a, \alpha_b ] = 2i\varepsilon_{abc} \Sigma_c,
\end{equation}
where $\varepsilon_{abc}$ is the Levi--Civita symbol.  
Then, we obtain 
\begin{gather}
\mathcal{O}^2 = \bm{\Pi}\cdot\bm{\Pi} + e\bm{B}\cdot\bm{\Sigma}, \\
[\mathcal{O},[\mathcal{O},-e\Phi] + i\dot{\mathcal{O}}] = -e (\nabla\cdot\bm{E}) - e [ \bm{\Pi}\cdot(\bm{\Sigma}\times\bm{E}) + (\bm{\Sigma}\times\bm{E})\cdot\bm{\Pi} ], \\
\begin{aligned}[b]
[\mathcal{O},[\mathcal{O},H']] &= \mathcal{O}^2 H' + H' \mathcal{O}^2 -2 \mathcal{O} H' \mathcal{O} \\
&= (\bm{\Pi}\cdot\bm{\Pi} + e\bm{B}\cdot\bm{\Sigma}) H' + H' (\bm{\Pi}\cdot\bm{\Pi} + e\bm{B}\cdot\bm{\Sigma}) - 2 (\bm{\alpha}\cdot\bm{\Pi}) H' (\bm{\alpha}\cdot\bm{\Pi}).  
\end{aligned}
\end{gather}
The last term apparently contains $\bm{\alpha}$ but can be eliminated; the explicit form depends on the commutation relation between $\bm{\alpha}$ and $H'$.  Therefore, we obtain the formal expression of the effective Hamiltonian as 
\begin{align}
H_\text{eff} &= m\beta + \mathcal{E}_4 + O(m^{-4}) \nonumber\\
&= m\beta - e\Phi + H' + \frac{1}{2m}\beta (\bm{\Pi}\cdot\bm{\Pi} + e\bm{B}\cdot\bm{\Sigma}) \nonumber\\
&\quad + \frac{e}{8m^2} (\nabla\cdot\bm{E}) + \frac{e}{8m^2} [ \bm{\Pi}\cdot(\bm{\Sigma}\times\bm{E}) + (\bm{\Sigma}\times\bm{E})\cdot\bm{\Pi} ] \nonumber\\
&\quad - \frac{1}{8m^2} [ (\bm{\Pi}\cdot\bm{\Pi} + e\bm{B}\cdot\bm{\Sigma}) H' + H' (\bm{\Pi}\cdot\bm{\Pi} + e\bm{B}\cdot\bm{\Sigma}) - 2 (\bm{\alpha}\cdot\bm{\Pi}) H' (\bm{\alpha}\cdot\bm{\Pi}) ] \nonumber\\
&\quad - \frac{1}{8m^3} \beta (\bm{\Pi}\cdot\bm{\Pi} + e\bm{B}\cdot\bm{\Sigma})^2 \nonumber\\
&\quad - \frac{1}{8m^3} \beta \{ -e (\nabla\cdot\bm{E}) - e [ \bm{\Pi}\cdot(\bm{\Sigma}\times\bm{E}) + (\bm{\Sigma}\times\bm{E})\cdot\bm{\Pi} ] \nonumber\\ 
&\hspace{50pt} + (\bm{\Pi}\cdot\bm{\Pi} + e\bm{B}\cdot\bm{\Sigma}) H' + H' (\bm{\Pi}\cdot\bm{\Pi} + e\bm{B}\cdot\bm{\Sigma}) - 2 (\bm{\alpha}\cdot\bm{\Pi}) H' (\bm{\alpha}\cdot\bm{\Pi}) \}^2 + O(m^{-4}).  
\label{eq:S_effective_formal}
\end{align}

\section{Isotropic TI model}

For the topological insulators Bi$_2$Se$_3$ and Bi$_2$Te$_3$, the $\bm{k}\cdot\bm{p}$ expansion to linear order in momentum $\bm{k}$ around the $\Gamma$ point becomes the Dirac Hamiltonian \cite{S_Zhang}
\begin{equation}
H(\bm{k}) = m \tau_z + A_2 (k_x \sigma_x + k_y \sigma_y) \tau_x + A_1 k_z \sigma_z \tau_x. 
\end{equation}
$\bm{\sigma}$ and $\bm{\tau}$ are the Pauli matrices for the spin and orbital degrees of freedom, respectively.  $\tau_z = \pm 1$ corresponds to the cation and anion $p$ orbitals. 
The model satisfy time-reversal, inversion, and three-fold rotational symmetries $\mathcal{T}=i\sigma_y \mathcal{K}$, $\mathcal{P}=\tau_z$, $C_3=\exp(i\pi\sigma_z/3)$, respectively, where $\mathcal{K}$ denotes the complex conjugate operator.  

With the rescaling of the momentum $\bm{k}$, we can eliminate the coefficients $A_1$ and $A_2$.  In addition, we include the electromagnetic potential $(\Phi, \bm{A})$, corresponding to the electric field $\bm{E}=-\nabla\Phi-\dot{\bm{A}}$ and the magnetic field $\bm{B} = \nabla\times\bm{A}$.  The electromagnetic potential replaces the momentum operator $\bm{p} = -i\nabla$ with the gauge-invariant momentum operator $\bm{\Pi} = \bm{p}+e\bm{A}$ from minimal coupling.  Here, the charge of an electron is $-e$ $(e>0)$.  
As a result, the Dirac Hamiltonian becomes 
\begin{equation}
H = m \beta + \bm{\alpha}\cdot(\bm{p}+e\bm{A}) - e\Phi. 
\label{eq:S_Dirac}
\end{equation}
We do not explicitly write the chemical potential hereafter.  
$\beta$ and $\bm{\alpha}$ are the $4\times 4$ matrices satisfying the relations 
\begin{gather}
\{ \alpha_a, \alpha_b \} = 0 \quad (a \neq b), \quad 
\{ \alpha_a, \beta \} = 0, \quad
\alpha_a^2 = \beta^2 = I, 
\label{eq:S_anticommutation}
\end{gather}
where $I$ is the identity matrix.  
For the present model, $\bm{\alpha}$ and $\beta$ are 
\begin{equation}
\beta = \tau_z, \quad \alpha_a = \sigma_a \tau_x.  
\end{equation}

We suppose doping of magnetic impurities, which couple to the Dirac electrons via the exchange coupling.  Since the model consists of the two $p$ orbitals of different origins, the strength of the exchange coupling depends on the orbitals.  Thus, the magnetic order $\bm{M}(\bm{r},t)$ affects the Dirac electrons in the form \cite{S_Wakatsuki}
\begin{align}
H' &= -J \bm{M}(\bm{r},t) \cdot \bm{\sigma} - J'\tau_z \operatorname{sgn}(m) \bm{M}(\bm{r},t) \cdot \bm{\sigma} \nonumber\\
&= -J \bm{M}(\bm{r},t) \cdot \bm{\sigma} - J' \tilde{\beta} \bm{M}(\bm{r},t) \cdot \bm{\sigma}. 
\label{eq:S_exchange_1}
\end{align}
We include the sign of the mass in the term with $J'$.  Since the the matrix $\beta$ appears with the mass $m$, its eigenvalue $\tilde{\beta} = \pm 1$ signifies the conduction or valence band in the large mass limit.  
Therefore, the strength of the exchange coupling is $J+J'$ for the conduction band and $J-J'$ for the valence band.

\subsection{Effective Hamiltonian}

For the Dirac Hamiltonian $H$ Eq.~\eqref{eq:S_Dirac} with the exchange coupling $H'$ Eq.~\eqref{eq:S_exchange_1}, we perform the unitary transformation to diagonalize the Hamiltonian.  We confirm that the exchange coupling commutes with the matrix $\beta$: $[H', \beta] = 0$.  

From Eq.~\eqref{eq:S_effective_formal}, we obtain the effective Hamiltonian to order $m^{-3}$ as 
\begin{align}
&\quad \ H_\text{eff} \nonumber\\
&= m\beta + \mathcal{E}_4 + O(m^{-4}) \nonumber\\
&= m\beta -e\Phi - J\bm{M}\cdot\bm{\sigma} - J'\tilde{\beta}\bm{M}\cdot\bm{\sigma} + \frac{1}{2m}\beta(\bm{\Pi}\cdot\bm{\Pi}+e\bm{\sigma}\cdot\bm{B}) \nonumber\\
&\quad + \frac{e}{8m^2}(\nabla\cdot\bm{E}) + \frac{e}{8m^2}[\bm{\Pi}\cdot(\bm{\sigma}\times\bm{E})+(\bm{\sigma}\times\bm{E})\cdot\bm{\Pi}] \nonumber\\
&\quad +\frac{J}{8m^2} \{ (\bm{\Pi}\cdot\bm{\sigma})[-i\nabla\cdot\bm{M} + \bm{\sigma}\cdot(\nabla\times\bm{M}) - 2i(\bm{\sigma}\times\bm{M})\cdot\bm{\Pi}] 
 + [i\nabla\cdot\bm{M} + \bm{\sigma}\cdot(\nabla\times\bm{M}) +2i\bm{\Pi}\cdot(\bm{\sigma}\times\bm{M})](\bm{\sigma}\cdot\bm{\Pi}) \} \nonumber\\
&\quad +\frac{J'\tilde{\beta}}{8m^2} \{ (\bm{\Pi}\cdot\bm{\sigma})[-i\nabla\cdot\bm{M} + \bm{\sigma}\cdot(\nabla\times\bm{M}) +2\bm{M}\cdot\bm{\Pi}] 
 + [i\nabla\cdot\bm{M} + \bm{\sigma}\cdot(\nabla\times\bm{M}) +2\bm{\Pi}\cdot\bm{M}](\bm{\sigma}\cdot\bm{\Pi}) \} \nonumber\\
&\quad -\frac{1}{8m^3}\beta(\bm{\Pi}\cdot\bm{\Pi}+e\bm{\sigma}\cdot\bm{B})^2 + \frac{1}{8m^3}\beta(\mathcal{F}^\dagger\cdot\mathcal{F}) + O(m^{-4}), 
\label{eq:S_effectiveH}
\end{align}
where the operator $\mathcal{F}$ is 
\begin{align}
\mathcal{F} = e\bm{\sigma}\cdot\bm{E} - J [\nabla\cdot\bm{M} + \bm{\Pi}\cdot(\bm{\sigma}\times\bm{M}) + (\bm{\sigma}\times\bm{M})\cdot\bm{\Pi}] - iJ'\tilde{\beta} [\bm{\sigma}\cdot(\nabla\times\bm{M}) + (\bm{\Pi}\cdot\bm{M}+\bm{M}\cdot\bm{\Pi})]. 
\end{align}
Since the effective Hamiltonian is diagonal in the orbital space, $\tilde{\beta}$ is regarded as an eigenvalue $\pm 1$ hereafter. 
As the present model is isotropic in the orbital space, we find $\bm{\Sigma} = \bm{\sigma}$, which corresponds to the spin of an electron. 
In the following, we consider the effective Hamiltonian to order $m^{-2}$.

\section{Response functions}

We consider the response functions in the metallic state using functional derivatives.  We first define the action $S$ using the effective Hamiltonian 
\begin{equation}
S = \int_k \bar{\psi}(k) [-i\omega_n + H_\text{eff}(k)] \psi(k)  
\end{equation}
with $k=(\bm{k},i\omega_n)$. 
For clarity, we use the simplified notation 
\begin{equation}
\int_k = T\sum_{\omega_n} \int_{\bm{k}} = T\sum_{\omega_n} \int \frac{d\bm{k}}{(2\pi)^3}
\end{equation}
where $T$ is the temperature and $\omega_n = (2n+1)\pi T$ ($n$: integer) is the fermionic Matsubara frequency.  The partition function $Z$ is given by the path integral 
\begin{equation}
Z = \int D\bar{\psi} D\psi e^{-S}.  
\end{equation}

Using a functional derivative of the partition function, we can calculate the expectation value of an operator $\hat{X}$.  The operator $\hat{X}$ should appear in the Hamiltonian in the form 
\begin{equation}
H_X = - \int d\bm{r} F_X(\bm{r},t) \hat{X}(\bm{r},t),
\end{equation}
where $F_X(\bm{r},t)$ is regarded as a generalized force that drives the quantity $X$.  Then, the expectation value $\langle \hat{X} \rangle$ satisfies 
\begin{equation}
\langle \hat{X}(\bm{r},t) \rangle = \frac{\delta \ln Z}{\delta F_X(\bm{r},t)} \bigg|_{F_X=0}.  
\end{equation}
Here, $\langle \ \rangle$ denotes the statistical average in the equilibrium.  
By further expanding the right-hand side, we can extract the effect of perturbations.  The linear response of $\langle \hat{X} \rangle$ to the generalized force $F_Y(\bm{r},t)$ becomes 
\begin{equation}
\langle \hat{X}(\bm{r},t) \rangle = \frac{\delta^2 \ln Z}{\delta F_X(\bm{r},t) \delta F_Y(\bm{r}',t')} \bigg|_{F_X=F_Y=0} F_Y(\bm{r}',t').  
\end{equation}
The result is equivalent to the Kubo formula.  We find that the coefficient of the linear response becomes the correlation function 
\begin{equation}
\frac{\delta^2 \ln Z}{\delta F_X(\bm{r},t) \delta F_Y(\bm{r}',t')} \bigg|_{F_X=F_Y=0} = \langle \hat{X}(\bm{r},t) \hat{Y}(\bm{r}',t') \rangle. 
\end{equation}
This expression hold when the expectation value vanishes in the equilibrium.  One can formally extend the expansion to higher orders of perturbations.  

The operator $\hat{X}$ has the form $\hat{X}(\bm{r},t) = \bar{\psi}(\bm{r},t) X(\bm{r},t) \psi(\bm{r},t)$, where $X(\bm{r},t)$ corresponds to the matrix representation of the operator.  Then, the calculation of the expectation value becomes the calculation of the connected diagrams of the Green's function.  We define the Green's function with the unperturbed Hamiltonian as 
\begin{equation}
G_0(\bm{k},i\omega_n) = -\langle \psi(\bm{k},i\omega_n) \bar{\psi}(\bm{k},i\omega_n) \rangle_0 = \frac{1}{i\omega_n - H_0(\bm{k}) + \mu - \Sigma(\omega_n)}. 
\label{eq:S_Green}
\end{equation}
Here, $\langle \ \rangle_0$ denotes the statistical average with the unperturbed Hamiltonian $H_0$.  The unperturbed Hamiltonian for the present system is 
\begin{equation}
H_0(\bm{k}) = m\beta + \frac{k^2}{2m} \beta, 
\end{equation}
and we use the empirical self-energy 
\begin{equation}
\Sigma = -\frac{i}{2\tau} \operatorname{sgn}(\omega_n)
\end{equation}
to describe diffusive transport with the lifetime $\tau$.  
To simplify the notation, we introduce the vertex function $\Gamma_X$ as 
\begin{equation}
-\frac{\delta S}{\delta F_X(q)}\bigg|_{F_X=0} = \int_k \bar{\psi}(k+q) \Gamma_{F_X}(k;q) \psi(k)
\end{equation}
with $q=(\bm{q},i\Omega_m)$ and the bosonic Matsubara frequency $\Omega_m = 2m\pi T$ ($m$: integer).  
Using Wick's theorem, we find that the correlation function becomes 
\begin{equation}
\langle \hat{X}(-q) \hat{Y}(q) \rangle_0
= -\operatorname{tr} \int_k \Gamma_{F_X}(k;-q) G_0(k+q) \Gamma_{F_Y}(k;q) G_0(k), 
\end{equation}
where $\operatorname{tr}$ stands for the trace of the matrix structure of the Hamiltonian; i.e., the spin matrices from the effective Hamiltonian for the present case.

\subsection{Vertex functions}

The analytic continuation requires the Matsubara frequency $i\Omega_m$ to be replaced with $\omega+i0^+$.  In the current model, bosonic Matsubara frequencies correspond to frequencies of external fields.  After the analytic continuation $i\Omega_m \to \omega+i0^+$, the vertex functions that we use below are 
\begin{gather}
\label{eq:S_vertex-i}
\Gamma^A_a(k;q) = -\frac{e\tilde{\beta}}{2|m|} (2k_a+q_a) - \frac{ie}{8m^2}(iq_0) q_a - \frac{e}{8m^2}(iq_0) \varepsilon_{abc}(2k_b+q_b)\sigma_c, \\
\Gamma^{A^2}_{ab}(k;q_1,q_2) = -\frac{e^2\tilde{\beta}}{|m|} - \frac{e^2}{4m^2} (iq_{1,0}-iq_{2,0}) \varepsilon_{abc} \sigma_c , \\
\Gamma^M_a(k;q) = (J+J'\tilde{\beta}) \sigma_a, \\
\begin{aligned}[b]
\Gamma^{AM}_{ab}(k;q,Q) &= -\frac{eJ}{4m^2}\left[ i\varepsilon_{abc}(Q_k+2q_k) + (\varepsilon_{adc}\varepsilon_{ceb}+\varepsilon_{abc}\varepsilon_{ced})\sigma_d(2k_e+q_e+Q_e) \right] \\
&\quad -\frac{eJ'\tilde{\beta}}{4m^2} \left[ -i\varepsilon_{abc}Q_c + (\delta_{ad}\delta_{be}+\delta_{ab}\delta_{de})\sigma_d(2k_e+q_e+Q_e) \right], 
\end{aligned}\\
\Gamma^{A^2M}_{abc}(k;q_1,q_2,Q) = -\frac{e^2 J}{4m^2} \left[ i\varepsilon_{bcd}(\sigma_a\sigma_d - \sigma_d\sigma_a) + i\varepsilon_{acd}(\sigma_b\sigma_d - \sigma_d\sigma_b) \right]
- \frac{e^2 J'\tilde{\beta}}{2m^2} (\sigma_a\delta_{bc} + \sigma_b\delta_{ac}).  
\label{eq:S_vertex-f}
\end{gather}
Here, $q_0$ refers to the temporal component of $q$ as a four-vector, i.e., $q_0 = \omega$ with $q = (\bm{q}, \omega)$.  
$\Gamma^{A^2}_{ab}(k;q_1,q_2)$ and $\Gamma^{AM}_{ab}(k;q,Q)$ are defined from the second-order derivatives of $-S$:  
\begin{gather}
-\frac{\delta^2 S}{\delta A_a(q_1) \delta A_b(q_2)} = \int_k \bar{\psi}(k+q_1+q_2) \Gamma^{A^2}_{ab}(k;q_1,q_2) \psi(k), \\
-\frac{\delta^2 S}{\delta A_a(q) \delta M_b(Q)} = \int_k \bar{\psi}(k+q+Q) \Gamma^{AM}_{ab}(k;q,Q) \psi(k). 
\end{gather}
Similarly, $\Gamma^{A^2M}_{abc}(k;q_1,q_2,Q)$ is given by
\begin{equation}
-\frac{\delta^3 S}{\delta A_a(q_1) \delta A_b(q_2) \delta M_c(Q)} = \int_k \bar{\psi}(k+q_1+q_2+Q) \Gamma^{A^2M}_{abc}(k;q_1,q_2,Q) \psi(k).  
\end{equation}
We hereafter omit the variables of $\Gamma^{M}_a$ as it is constant.

\subsection{Current operator}

We derive the current operator here.  Though we do not use it in the following calculations, it is worth knowing that the magnetic order affects the charge current.  
The functional derivative of the action gives the current operator $\hat{\bm{\mathcal{J}}}$ as 
\begin{align}
\hat{\bm{\mathcal{J}}}(\bm{r},t) &= -\frac{\delta S}{\delta \bm{A}(\bm{r},t)} 
= \bar{\psi}(\bm{r},t) \bm{\mathcal{J}}(\bm{r},t) \psi(\bm{r},t), 
\label{eq:S_current_definition}
\end{align}
where $\bm{\mathcal{J}}(\bm{r},t)$ becomes 
\begin{align}
\bm{\mathcal{J}}
=& -\frac{e}{m} \beta \bm{\Pi} -\frac{e^2}{4m^2} \bm{\sigma}\times\bm{E} \nonumber\\
& -\frac{eJ}{4m^2} \Big[ \nabla\times\bm{M} -2i(\bm{M}\times\bm{\Pi}+\bm{\Pi}\times\bm{M}) 
- \bm{\sigma}\times(\bm{M}\times\bm{\Pi}) - (\bm{\Pi}\times\bm{M})\times\bm{\sigma} 
- \bm{M}\times(\bm{\sigma}\times\bm{\Pi}) - (\bm{\Pi}\times\bm{\sigma})\times\bm{M} \Big] \nonumber\\
& -\frac{eJ'\tilde{\beta}}{4m^2} \Big[ \nabla\times\bm{M}
+ \bm{\sigma}(\bm{M}\cdot\bm{\Pi}) + (\bm{\Pi}\cdot\bm{M})\bm{\sigma} +\bm{M}(\bm{\sigma}\cdot\bm{\Pi}) + (\bm{\Pi}\cdot\bm{\sigma})\bm{M} \Big] + O(m^{-3}). 
\end{align}
The result corresponds to the sum of $\Gamma^A$, $\Gamma^{A^2}$, $\Gamma^{AM}$, and $\Gamma^{A^2M}$.  
Alternatively, we can obtain $\bm{\mathcal{J}}$ at $\bm{A}=\bm{0}$ from the relation $\bm{\mathcal{J}} = ie[\bm{r},H_\text{eff}]$ without inserting the electromagnetic potential.

\section{Response in the metallic state}

In this section, we show the detailed calculations of various response functions that we presented in the main part. 
The density of states (DOS) repeatedly appears in the following calculations.  For the unperturbed Hamiltonian $H_0$, the DOS at the chemical potential $\mu$ is 
\begin{align}
D_0(\mu) &= 2\sum_{\tilde{\beta}} \int_{\bm{k}} \delta \left( \mu - \tilde{\beta} \left(|m|+\frac{k^2}{2|m|}\right) \right) 
= \frac{\sqrt{2}}{\pi^2} |m|^{3/2} \left( |\mu|-|m| \right)^{1/2} \Theta\left( |\mu|-|m| \right),
\end{align}
where we use the step function 
\begin{equation}
\Theta(x) = 
\begin{cases}
1 & (x \geq 0), \\
0 & (x < 0).
\end{cases}
\end{equation}
When we discuss the metallic state, it is convenient to measure the chemical potential from the band edge, which we denote as $\epsilon$: 
\begin{equation}
\epsilon_F = \operatorname{sgn}(\mu) (|\mu|-|m|) \quad \text{for } |\mu| \geq |m|. 
\end{equation}
Then, the carrier density is 
\begin{equation}
n(\epsilon_F) 
= \int_0^\mu d\mu' D_0(\mu') 
= \frac{2\sqrt{2}}{3\pi^2} \operatorname{sgn}(\epsilon_F)|m|^{3/2}  |\epsilon_F|^{3/2}. 
\label{eq:S_density}
\end{equation}

The DOS and the carrier density are related by 
\begin{equation}
D_0(\mu) = \frac{3n(\epsilon_F)}{2\epsilon_F}. 
\end{equation}

\subsection{Current response under an electric field}

\subsubsection{Linear response}

Using the partition function, we can write the linear current response to the external electric field as 
\begin{equation}
j_a(\omega) = \frac{1}{i\omega} \frac{\delta^2 \ln Z}{\delta A_a(-\omega) \delta A_b(\omega)} \bigg|_{A=0} E_b(\omega). 
\end{equation}
We note that this equation is equivalent to the Kubo formula.  The coefficient corresponds to the electric conductivity: 
\begin{equation}
\sigma_{ab}(\omega) = \frac{1}{i\omega} \frac{\delta^2 \ln Z}{\delta A_a(-\omega) \delta A_b(\omega)} \bigg|_{A=0}.  
\end{equation}
To one-loop order, the conductivity becomes 
\begin{align}
\sigma_{ab}(\omega) &= -\frac{1}{i\omega} \operatorname{tr} \int_k \Gamma^A_a(k;-q) G_0(k+q) \Gamma^A_b(k;q) G_0(k) 
+ \frac{1}{i\omega} \operatorname{tr} \int_k \Gamma^{A^2}_{ab}(k;-q,q) G_0(k) \nonumber\\
&= \frac{e^2}{|m|} |n(\epsilon)| \frac{\tau}{1-i\omega\tau} \delta_{ab}. 
\label{eq:S_conductivity}
\end{align}
The result coincides with the one obtained from the Drude model.

\subsubsection{Effect of a magnetic order without a uniform magnetization}

Next, we consider the effect of the magnetic order, which modifies the conductivity.  As we assume that the magnetic order does not have uniform magnetization and is characterized by finite wavevectors, the lowest-order correction by the magnetic order appears at second order.  Therefore, we can write the correction to the conductivity by the magnetic order as  
\begin{equation}
j_a(\omega) = \frac{1}{2} \sum_{\bm{Q}} \eta_{abcd}(\omega,\bm{Q}) E_b(\omega) M_c(\bm{Q}) M_d(-\bm{Q}), 
\end{equation}
where the coefficient is given by 
\begin{equation}
\eta_{abcd}(\omega) = \frac{1}{i\omega} \frac{\delta^4 \ln Z}{\delta A_a(-\omega) \delta A_b(\omega) \delta M_c(\bm{Q}) \delta M_d(-\bm{Q})}. 
\end{equation}
Using the vertex functions Eqs.~\eqref{eq:S_vertex-i}--\eqref{eq:S_vertex-f}, we obtain the following five contributions: 
\begin{gather}
\begin{aligned}[b]
\eta_{abcd}^{(1)}(\omega,\bm{Q}) 
&=\frac{1}{i\omega} \left\langle \int_k \bar{\psi}(k+Q) \Gamma^{AM}_{ac}(k+q;-q,Q) \psi(k+q) \int_{k_1} \bar{\psi}(k_1+q) \Gamma^{AM}_{bd}(k_1+Q;q,-Q) \psi(k_1+Q) \right\rangle_{\!\!0} \\ 
&\quad + (c \leftrightarrow d, Q \to -Q), 
\end{aligned}\\
\begin{aligned}[b]
\eta_{abcd}^{(2)}(\omega,\bm{Q}) 
&= \frac{1}{i\omega} \bigg\langle \int_k \bar{\psi}(k+Q) \Gamma^{AM}_{ac}(k+q;-q,Q) \psi(k+q) \int_{k_1} \bar{\psi}(k_1+q) \Gamma^A_b(k_1;q) \psi(k_1) \int_{k_2} \bar{\psi}(k_2-Q) \Gamma^M_d \psi\mathrlap{(k_2) \bigg\rangle_{\!\!0}} \\ 
&\quad + (c \leftrightarrow d, Q \to -Q), 
\end{aligned}\\
\begin{aligned}[b]
\eta_{abcd}^{(3)}(\omega,\bm{Q}) 
&= \frac{1}{i\omega} \bigg\langle \int_k \bar{\psi}(k) \Gamma^A_a(k+q;-q) \psi(k+q) \int_{k_1} \bar{\psi}(k_1+q+Q) \Gamma^{AM}_{bc}(k_1;q,Q) \psi(k_1) \int_{k_2} \bar{\psi}(k_2-Q) \Gamma^M_d \psi\mathrlap{(k_2) \bigg\rangle_{\!\!0}} \\ 
&\quad + (c \leftrightarrow d, Q \to -Q), 
\end{aligned}\\
\eta_{abcd}^{(4)}(\omega,\bm{Q}) = \frac{1}{i\omega} \left\langle \int_k \bar{\psi}(k+Q) \Gamma^{A^2M}_{abc}(k;-q,q,Q) \psi(k) \int_{k_1} \bar{\psi}(k_1) \Gamma^M_d \psi(k_1+Q) \right\rangle_{\!\!0} + (c \leftrightarrow d, Q \to -Q), \\
\begin{aligned}[b]
&\quad\ \eta_{abcd}^{(5)}(\omega,\bm{Q}) \\
&= \frac{1}{i\omega} \bigg\langle \int_k \bar{\psi}(k) \Gamma^A_a(k+q;-q) \psi(k+q) \int_{k_1} \bar{\psi}(k_1+q) \Gamma^A_b(k_1;q) \psi(k_1) \int_{k_2} \bar{\psi}(k_2+Q) \Gamma^M_c \psi(k_2) \int_{k_3} \bar{\psi}(k_3-Q) \Gamma^M_d \psi(k_3) \bigg\rangle_{\!\!0}, 
\end{aligned}\\
\begin{aligned}[b]
\eta_{abcd}^{(6)}(\omega,\bm{Q}) 
&= \frac{1}{i\omega} \left\langle \int_k \bar{\psi}(k) \Gamma^{A^2}_{ab}(k;-q,q) \psi(k) \int_{k_1} \bar{\psi}(k_1+Q) \Gamma^M_c \psi(k_1) \int_{k_2} \bar{\psi}(k_2-Q) \Gamma^M_d \psi(k_2) \right\rangle_{\!\!0}. 
\end{aligned} 
\end{gather}
Here we use the notation $q = (\bm{0},i\Omega_m)$ and $Q = (\bm{Q},0)$.  After the momentum integrations, the summation of Matsubara frequencies, and the analytic continuation, we obtain 
\begin{gather}
\begin{aligned}[b]
\eta^{(1)}_{abcd}(\omega,\bm{Q}) = \frac{e^2}{4|m|^3} |n(\epsilon_F)| \frac{\tau}{1-i\omega\tau} \{ &
J^2 [10\delta_{ab}\delta_{cd} - 3(\delta_{ac}\delta_{bd}+\delta_{ad}\delta_{bc})]
+ J'^2 [2\delta_{ab}\delta_{cd}+5(\delta_{ac}\delta_{bd}+\delta_{ad}\delta_{bd})] \\
& -2JJ'\tilde{\beta} [2\delta_{ab}\delta_{cd} + (\delta_{ac}\delta_{bd}+\delta_{ad}\delta_{bc})]
\}, 
\end{aligned}\\
\eta^{(5)}_{abcd}(\omega,\bm{Q}) = -\frac{4e^2}{|m|} |n(\epsilon_F)| (J+J'\tilde{\beta})^2 \frac{\tau^3}{(1-i\omega\tau)^3} \left( 1+\frac{\omega^2}{8m^2} \right) \delta_{ab} \delta_{cd}, \\
\eta^{(2)}_{abcd}(\omega,\bm{Q}) + \eta^{(3)}_{abcd}(\omega,\bm{Q}) = \eta^{(4)}_{abcd}(\omega,\bm{Q}) = \eta^{(6)}_{abcd}(\omega,\bm{Q}) = 0. 
\end{gather}

\subsubsection{Effect of a uniform magnetization}

When the local magnetic moments have a uniform magnetization $\bm{M}_{\bm{0}} \neq \bm{0}$, we expect the anomalous Hall effect $j_a(\omega) = \sigma^\text{AH}_{ab}(\omega) E_b(\omega)$ with the anomalous Hall conductivity $\sigma^\text{AH}_{abc}(\omega) \propto \varepsilon_{abc} M_c(\bm{0})$.  
One may calculate the anomalous Hall conductivity with the Green's function Eq.~\eqref{eq:S_Green} and the vertices Eqs.~\eqref{eq:S_vertex-i}--\eqref{eq:S_vertex-f}; however, one does not find the anomalous Hall conductivity at zero frequency $\sigma^\text{AH}_{ab}(0)$ with the same procedure.  
We need a nonperturbative effect to the model, i.e., a correction to the unperturbed Hamiltonian.  
Here, we define the unperturbed Hamiltonian as 
\begin{equation}
H_\text{mag}(\bm{k}) = |m|\tilde{\beta} + \frac{k^2}{2|m|}\tilde{\beta} - m_z \sigma_z.  
\end{equation}
For simplicity but without loss of generality, we assume that the uniform magnetization is oriented along the $z$ axis.  $m_z$ represents the spin polarization on the Fermi surface, arising from the exchange coupling $-(J+J'\tilde{\beta})\bm{M}\cdot\bm{\sigma}$: 
\begin{equation}
m_z \approx (J+J'\tilde{\beta}) M_z(\bm{0}). 
\end{equation}
The unperturbed Green's function uses $H_\text{mag}$ instead of $H_0$: 
\begin{equation}
G_{0,\text{mag}}(\bm{k},i\omega_n) = \frac{1}{i\omega_n - H_\text{mag}(\bm{k}) + \mu - \Sigma(\omega_n)}. 
\end{equation}
Now we can calculate the anomalous Hall conductivity similarly as Eq.~\eqref{eq:S_conductivity} with $G_0$  replaced with $G_{0,\text{mag}}$.  For $\tau^{-1}, |m_z| \ll |\epsilon_F|$, we obtain the anomalous Hall conductivity at low frequencies $(|\omega| \ll \tau^{-1})$
\begin{equation}
\sigma_{ab}^\text{AH}(\omega) \approx \frac{e^2}{2m^2} D_0(\mu) \varepsilon_{abz} m_z 
= \frac{3e^2}{4m^2} \frac{n(\epsilon_F)}{\epsilon_F} \varepsilon_{abz} m_z.  
\end{equation}

Finite magnetization of the local magnetic moments forces spin polarization of the conduction electrons through the Zeeman coupling.  We can understand the spin polarization of the conduction electrons as the spin-dependent Fermi energies $\epsilon_F \pm m_z$.  The spin-polarized conduction electrons with the spin-orbital coupling inherent in the Dirac Hamiltonian lead to the finite anomalous Hall conductivity.

\subsection{Magnetization by current}

In the presence of an inversion-breaking magnetic order, the symmetry analysis allows finite uniform magnetization under an external electric field, i.e., electric current.  As we have discussed, the lowest-order contributions appear at order $M^2$, the induced uniform magnetization of Dirac electrons should have the form 
\begin{equation}
\langle \sigma_a \rangle(\omega) = \frac{1}{2} \sum_{\bm{Q}} \lambda_{abcd}(\omega,\bm{Q}) E_b(\omega) M_c(\bm{Q}) M_d(-\bm{Q}).  
\end{equation}
The coefficient $\lambda$ takes the form 
\begin{equation}
\lambda_{abcd}(\omega) = \frac{1}{i\omega} \frac{\delta^3}{\delta A_b(\omega) \delta M_c(\bm{Q}) \delta M_d(-\bm{Q})} \left\langle \int_k \bar{\psi}(k)\sigma_a\psi(k+q) \right\rangle, 
\end{equation}
from which we find the two contributions 
\begin{gather}
\begin{aligned}[b]
\lambda^{(1)}_{abcd}(\omega,\bm{Q}) 
&= \frac{1}{i\omega} \bigg\langle \int_k \bar{\psi}(k)\sigma_a\psi(k+q) \int_{k_1} \bar{\psi}(k_1+Q+q) \Gamma^{AM}_{bc}(k_1;q,Q) \psi(k_1) \int_{k_2} \bar{\psi}(k_2-Q) \Gamma^M_d \psi(k_2) \mathrlap{\bigg\rangle_{\!\!0}} \\
&\quad + (c \leftrightarrow d, Q \to -Q), 
\end{aligned}\\
\begin{aligned}[b]
\lambda^{(2)}_{abcd}(\omega,\bm{Q}) 
&= \frac{1}{i\omega} \bigg\langle \int_k \bar{\psi}(k)\sigma_a\psi(k+q) \int_{k_1} \bar{\psi}(k_1+q) \Gamma^{A}_b(k_1;q) \psi(k_1) \int_{k_2} \bar{\psi}(k_2+Q) \Gamma^M_c \psi(k_2) \int_{k_3} \bar{\psi}(k_3-Q) \Gamma^M_d \psi(k_3) \bigg\rangle_{\!\!0},
\end{aligned}
\end{gather}
with $q = (\bm{0},i\Omega_m)$ and $Q = (\bm{Q},0)$.  
Using Wick's theorem, we can calculate the two contributions to obtain 
\begin{gather}
\begin{aligned}[b]
\lambda^{(1)}_{abcd}(\omega,\bm{Q}) &= \frac{1}{i\omega} \operatorname{tr} \int_k \sigma_a G_0(k+q) \Gamma^{AM}_{bc}(k;q,Q) G_0(k-Q) \Gamma^M_d G_0(k) \\
&\quad +\!\frac{1}{i\omega} \operatorname{tr} \int_k \sigma_a G_0(k+q) \Gamma^M_d G_0(k+Q+q) \Gamma^{AM}_{bc}(k;q,Q) G_0(k) \\
&\quad + (c \leftrightarrow d, Q \to -Q) \\
&= -\frac{ie}{m^2} n(\epsilon_F) (J+J'\tilde{\beta}) \frac{\tau^3}{(1-i\omega\tau)^3}
[ J(-4\varepsilon_{acd}Q_b - \varepsilon_{ade}Q_e\delta_{bc} + \varepsilon_{ace}Q_m\delta_{bd} + \varepsilon_{abd}Q_c - \varepsilon_{abc}Q_d) \\
&\hspace{140pt} + J'\tilde{\beta} (\varepsilon_{ade}Q_e\delta_{bc} - \varepsilon_{ace}Q_e\delta_{bd} -\varepsilon_{abd}Q_c + \varepsilon_{abc}Q_d) ] + O(Q^2),
\end{aligned}\\
\begin{aligned}[b]
\lambda^{(2)}_{abcd}(\omega,\bm{Q}) 
&= -\frac{1}{i\omega} \operatorname{tr} \int_k \sigma_a G_0(k+q) \Gamma^A_b(k;q) G_0(k) \Gamma^M_c G_0(k-Q) \Gamma^M_d G_0(k) \\
&\quad -\!\frac{1}{i\omega} \operatorname{tr} \int_k \sigma_a G_0(k+q) \Gamma^M_c G_0(k+q-Q) \Gamma^M_d G_0(k+q) \Gamma^A_b(k;q) G_0(k) \\
&\quad -\!\frac{1}{i\omega} \operatorname{tr} \int_k \sigma_a G_0(k+q) \Gamma^M_c G_0(k+q-Q) \Gamma^A_b(k-Q;q) G_0(k-Q) \Gamma^M_d G_0(k) \\
&\quad + (c \leftrightarrow d, Q \to -Q) \\
&= -\frac{e}{m^2} n(\epsilon_F) (J+J'\tilde{\beta})^2 \frac{\omega\tau^4}{(1-i\omega\tau)^4} (\delta_{ac}\varepsilon_{bed}-\delta_{ad}\varepsilon_{bec}) Q_e + O(Q^2).
\end{aligned}
\end{gather}
Here, we should expand the coefficient $\lambda$ with respect to the wavevector $\bm{Q}$ and extract odd-order contributions to capture inversion breaking of the magnetic order.  In the results above, we retain the terms to linear order in $Q$.  The second term $\lambda^{(2)}$ is proportional to $\omega$, so that it does not contribute to static uniform magnetization.

\subsection{Current by an external magnetic field}

An oscillating external magnetic field may induce electric current if the system breaks inversion.  As the model that we consider here does not break inversion without a magnetic order, an inversion-breaking magnetic order is necessary for current response.  The uniform current response should have the form 
\begin{equation}
j_a(\omega) = \frac{1}{2} \sum_{\bm{Q}} \kappa_{abcd}(\omega,\bm{Q}) B_b(\omega) M_d(\bm{Q}) M_d(-\bm{Q}).  
\end{equation}
In theory, it is convenient to consider 
\begin{equation}
j_a(\omega) = \frac{1}{2} \sum_{\bm{Q}} \tilde{\kappa}_{a\tilde{b}cd}(\omega,\bm{q},\bm{Q}) A_{\tilde{b}}(\bm{q},\omega) M_c(\bm{Q}) M_d(-\bm{Q}).  
\end{equation} 
Since the uniform magnetic field and the vector potential are related by $\bm{B}(\omega) = i\bm{q}\times\bm{A}(\bm{q},\omega)$, we should expand $\tilde{\kappa}$ with respect to $\bm{q}$ to find 
\begin{equation}
\label{eq:S_gauge-invariance}
\tilde{\kappa}_{a\tilde{b}cd}(\omega,\bm{q},\bm{Q}) = \kappa_{abcd}(\omega,\bm{Q}) \cdot i\varepsilon_{b\tilde{a}\tilde{b}} q_{\tilde{a}}. 
\end{equation}
We can calculate $\tilde{\kappa}$ from 
\begin{equation}
\tilde{\kappa}_{a\tilde{b}cd}(\omega,\bm{q},\bm{Q}) = \frac{\delta^4 \ln Z}{\delta A_a(-\omega) \delta A_{\tilde{b}}(\bm{q},\omega) \delta M_c(\bm{Q}) \delta M_d(-\bm{Q})}.  
\end{equation}
Using the vertex functions Eqs.~\eqref{eq:S_vertex-i}--\eqref{eq:S_vertex-f}, we find the five contributions 
\begin{gather}
\begin{aligned}[b]
\tilde{\kappa}_{a\tilde{b}cd}^{(1)}(\omega,\bm{q},\bm{Q}) 
&= \bigg\langle \int_k \bar{\psi}(k+Q) \Gamma^{AM}_{ac}(k+q;-q,Q) \psi(k+q) \int_{k_1} \bar{\psi}(k_1+q) \Gamma^{AM}_{\tilde{b}d}(k_1+Q;q,-Q) \psi(k_1+Q) \bigg\rangle_{\!\!0} \\ 
&\quad + (c \leftrightarrow d, Q \to -Q), 
\end{aligned}\\
\begin{aligned}[b]
\tilde{\kappa}_{a\tilde{b}cd}^{(2)}(\omega,\bm{q},\bm{Q}) 
&= \bigg\langle \int_k \bar{\psi}(k+Q) \Gamma^{AM}_{ac}(k+q;-q,Q) \psi(k+q) \int_{k_1} \bar{\psi}(k_1+q) \Gamma^A_{\tilde{b}}(k_1;q) \psi(k_1) \int_{k_2} \bar{\psi}(k_2-Q) \Gamma^M_d \psi\mathrlap{(k_2) \bigg\rangle_{\!\!0}} \\ 
&\quad + (c \leftrightarrow d, Q \to -Q), 
\end{aligned}\\
\begin{aligned}[b]
\tilde{\kappa}_{a\tilde{b}cd}^{(3)}(\omega,\bm{q},\bm{Q}) 
&= \bigg\langle \int_k \bar{\psi}(k) \Gamma^A_a(k+q;-q) \psi(k+q) \int_{k_1} \bar{\psi}(k_1+q+Q) \Gamma^{AM}_{\tilde{b}c}(k_1;q,Q) \psi(k_1) \int_{k_2} \bar{\psi}(k_2-Q) \Gamma^M_d \psi\mathrlap{(k_2) \bigg\rangle_{\!\!0}} \\ 
&\quad + (c \leftrightarrow d, Q \to -Q), 
\end{aligned}\\
\tilde{\kappa}_{abcd}^{(4)}(\omega,\bm{q},\bm{Q}) = \left\langle \int_k \bar{\psi}(k+Q) \Gamma^{A^2M}_{a\tilde{b}c}(k;-q,q,Q) \psi(k) \int_{k_1} \bar{\psi}(k_1) \Gamma^M_d \psi(k_1+Q) \right\rangle_{\!\!0} + (c \leftrightarrow d, Q \to -Q), \\
\begin{aligned}[b]
&\quad\ \tilde{\kappa}_{a\tilde{b}cd}^{(5)}(\omega,\bm{q},\bm{Q}) \\
&= \bigg\langle \int_k \bar{\psi}(k) \Gamma^A_a(k+q;-q) \psi(k+q) \int_{k_1} \bar{\psi}(k_1+q) \Gamma^A_{\tilde{b}}(k_1;q) \psi(k_1) \int_{k_2} \bar{\psi}(k_2+Q) \Gamma^M_c \psi(k_2) \int_{k_3} \bar{\psi}(k_3-Q) \Gamma^M_d \psi(k_3) \bigg\rangle_{\!\!0},
\end{aligned}\\
\begin{aligned}[b]
\tilde{\kappa}_{a\tilde{b}cd}^{(6)}(\omega,\bm{q},\bm{Q}) 
&= \bigg\langle \int_k \bar{\psi}(k) \Gamma^{A^2}_{a\tilde{b}}(k;-q,q) \psi(k) \int_{k_1} \bar{\psi}(k_1+Q) \Gamma^M_c \psi(k_1) \int_{k_2} \bar{\psi}(k_2-Q) \Gamma^M_d \psi(k_2) \bigg\rangle_{\!\!0}, 
\end{aligned}
\end{gather}
with $q = (\bm{q},i\Omega_m)$ and $Q = (\bm{Q},0)$.  
By evaluating the expressions, we can see that $\kappa_{abcd}$ vanishes at least to order $Q J^2 n(\epsilon)/m^2$; therefore, we neglect the current directly induced by an oscillating external magnetic field.  We cannot exclude the possibility of finite contributions at order $n(\epsilon) J^2/m^4$ or $n(\epsilon) J^2/(m^3 (|\mu|-|m|))$ here.  We emphasize that careful calculations are necessary, which should satisfy the gauge invariance Eq.~\eqref{eq:S_gauge-invariance}.

\subsection{Current response by an oscillating magnetic order}
\label{sec:S_current-response}

As we have observed that the uniform current directly induced by an external oscillating magnetic field is negligible, we then investigate the current induced by an oscillation of the magnetic order, which has the form 
\begin{equation}
j_a(\omega) = \frac{1}{2} \sum_{\substack{\bm{Q} \\ \omega_1+\omega_2=\omega}} \gamma_{abc}(\omega_1,\omega_2,\bm{Q}) M_b(\bm{Q},\omega_1) M_c(-\bm{Q},\omega_2),
\end{equation}
where the coefficient is given by 
\begin{equation}
\gamma_{abc}(\omega_1,\omega_2,\bm{Q}) = \frac{\delta^3 \ln Z}{\delta A_a(-\omega_1-\omega_2) \delta M_b(\bm{Q},\omega_1) \delta M_c(-\bm{Q},\omega_2)}. 
\end{equation}
$\gamma$ has the two distinct contributions 
\begin{gather}
\begin{aligned}[b]
\gamma_{abc}^{(1)}(\omega_1,\omega_2,\bm{Q})  
&= \left\langle \int_k \bar{\psi}(k) \Gamma^A_a(k+q;-q) \psi(k+q) \int_{k_1} \bar{\psi}(k_1+Q_1) \Gamma^M_b \psi(k_1) \int_{k_2} \bar{\psi}(k_2+Q_2) \Gamma^M_c \psi(k_2) \right\rangle_{\!\!0},
\end{aligned}\\
\begin{aligned}[b]
\gamma_{abc}^{(2)}(\omega_1,\omega_2,\bm{Q}) 
= \left\langle \int_k \bar{\psi}(k+Q_1) \Gamma^{AM}_{ab}(k+q;-q,Q_1) \psi(k+q) \int_{k_1} \bar{\psi}(k_1+Q_2) \Gamma^M_c \psi(k_1) \right\rangle_{\!\!0} + (b \leftrightarrow c, Q_1 \leftrightarrow Q_2),
\end{aligned}
\end{gather}
where we use the notations $q = (\bm{0},i\Omega_m)$, $Q_1 = (\bm{Q},i\Omega_{m_1})$, and $Q_2 = (-\bm{Q},i\Omega_{m_2})$ with the analytic continuations $i\Omega_m \to \omega + i0^+$, $i\Omega_{m_1} \to \omega_1 + i0^+$, and $i\Omega_{m_2} \to \omega + i0^+$.  
Using Wick's theorem, we can calculate the expressions to obtain 
\begin{gather}
\begin{aligned}[b]
\gamma^{(1)}_{abc}(\omega_1,\omega_2,\bm{Q}) 
&= \operatorname{tr} \int_k \Gamma^A_a(k+q;-q) G_0(k+q) \Gamma^M_b G_0(k+Q_2) \Gamma^M_c G_0(k) + (b \leftrightarrow c, Q_1 \leftrightarrow Q_2) \\
&= \frac{e}{4m^2} n(\epsilon_F) (J+J'\tilde{\beta})^2 \frac{\omega_1+\omega_2}{\omega_1+\omega_2+i/\tau} \left[ \frac{\omega_1}{(\omega_1+i/\tau)^2} + \frac{\omega_2}{(\omega_2+i/\tau)^2} \right] \varepsilon_{ade} Q_d \varepsilon_{ebd} + O(Q^2),  
\end{aligned}\\
\begin{aligned}[b]
\gamma^{(2)}_{abc}(\omega_1,\omega_2,\bm{Q})
&= -\operatorname{tr} \int_k \Gamma^{AM}_{ab}(k+Q_2;-q,Q_1) G_0(k+Q_2) \Gamma^M_c G_0(k) + (b \leftrightarrow c, Q_1 \leftrightarrow Q_2) \\
&= -\frac{e}{2m^2} n(\epsilon_F) (J+J'\tilde{\beta}) 
\left[ \frac{\omega_2}{(\omega_2+i/\tau)^2} - \frac{\omega_1}{(\omega_1+i/\tau)^2} \right] 
\\ &\qquad\times
\left[ J(\varepsilon_{acd}\varepsilon_{deb}+\varepsilon_{abd}\varepsilon_{dec}) + J'\tilde{\beta} (\delta_{ac}\delta_{be}+\delta_{ab}\delta_{ce}) \right] Q_e + O(Q^2). 
\end{aligned}
\end{gather}
Like the calculation of the magnetization induced by an external electric field, we should expand the coefficient with respect to the wavevector $\bm{Q}$ and extract the odd-order contributions in $\bm{Q}$ to capture inversion breaking by the magnetic order.  We can confirm that the uniform current vanishes in the zero-frequency limit $(\omega_1, \omega_2 \to 0)$ as there must be no uniform current in the equilibrium.

\section{Polarization}

This section deals with an insulating case, where the chemical potential $\mu$ lies inside the gap $(|\mu| < |m|)$.  We focus on the the orbital part of the polarization, which reflects the geometric properties of the wave function.  Without the magnetic order or the exchange coupling, the electronic system itself preserves inversion and hence there is no polarization.  An adiabatic insertion of the exchange coupling may develops finite polarization in the presence of an inversion-breaking magnetic order.  We can unambiguously quantify the polarization by measuring from the inversion-symmetric state.

To calculate the electric polarization in an insulator, we follow the method by King-Smith and Vanderbilt \cite{S_King-Smith}.  Suppose that the Hamiltonian $H_\lambda$ describes the electronic system, where the parameter $\lambda$ $(0 \leq \lambda \leq 1)$ continuously alters the potential.  They showed that the change in the polarization per unit volume by an adiabatic change of the parameter $\lambda$ is 
\begin{equation}
\Delta P_i = ie \sum_{\substack{n \\ \text{(occupied)}}} \int_\text{BZ} \frac{d^3k}{(2\pi)^3} \int_{0}^{1} d\lambda \langle \partial_{k_a} u_{n\bm{k}} | \partial_\lambda u_{n\bm{k}} \rangle + \text{c.c.}
\end{equation}
Here we write the Bloch wave function as $\psi_{n\bm{k}} = u_{n\bm{k}} e^{i\bm{k}\cdot\bm{r}}$ with the band index $n$ and the lattice-periodic part $u_{n\bm{k}}$, and the charge of an electron is $-e$ $(e>0)$.  The momentum integration is performed in the Brillouin zone and the band index is summed over the all occupied bands.  The polarization is defined modulo the lattice period.  
This expression is concise, but it contains the wave function, which makes an analytic calculation difficult.  There is an equivalent expression that uses the Green's function $G_\lambda = (\omega - H_\lambda)^{-1}$ to calculate the orbital-part of the polarization \cite{S_WangQiZhang,S_Lee}: 
\begin{align}
\Delta P_i 
&= -\frac{e}{2} \int \frac{d\omega}{2\pi} \int_\text{BZ} \frac{d^3k}{(2\pi)^3} \int_0^1 d\lambda [\operatorname{tr} (G_\lambda \partial_{k_i} G_\lambda^{-1} \partial_\lambda G_\lambda) - \operatorname{tr} (G_\lambda \partial_\lambda G_\lambda^{-1} \partial_{k_i} G_\lambda)]. 
\label{eq:S_polarization}
\end{align}
Here, the trace tr includes the summation over all bands.  

In the following, we consider the massive Dirac Hamiltonian 
\begin{equation}
H = m\beta + \bm{\alpha}\cdot\bm{k}
\end{equation}
as the unperturbed electronic Hamiltonian and the exchange coupling with the magnetic order 
\begin{equation}
H' = -J\bm{M}\cdot\bm{\sigma}
\end{equation}
as the source of polarization.  The Green's function becomes $G_J = (\omega - H - H')^{-1}$.  Note that the Dirac Hamiltonian is defined in the continuum without any lattice structure.  The magnetic order with the wavevector $\bm{Q}$ introduces the periodicity to the system, which could be interpreted as the Brillouin zone.  
In considering the polarization in an insulating state, we can set the chemical potential $\mu = 0$, so that it lies in the mass gap.  When the exchange coupling is much smaller than the Dirac mass $(|JM| \ll |m|)$, we can treat the exchange coupling as a perturbation.  Then, the expression of the polarization Eq.~\eqref{eq:S_polarization} has the integration over the Brillouin zone and the summation of the band index, which can be replaced by the integration of the momentum for $-\infty < k_a < \infty$.  Henceforth, the trace means the matrix trace originated from the $4 \times 4$ Hamiltonian.  As a result, the formula for the polarization becomes 
\begin{align}
\Delta P_a = -\frac{e}{2} \int \frac{d\omega}{2\pi} \int \frac{d^3k}{(2\pi)^3} \int_{0}^{J} d\tilde{J} [ \operatorname{tr}(G_{\tilde{J}}\partial_{k_a}G_{\tilde{J}}^{-1}\partial_{\tilde{J}}G_{\tilde{J}}) - \operatorname{tr}(G_{\tilde{J}}\partial_{\tilde{J}}G_{\tilde{J}}^{-1}\partial_{k_a}G_{\tilde{J}}) ].  
\end{align}
The polarization is measured from the state with $J=0$, where the electronic system remains centrosymmetric and hence the polarization vanishes.

The Green's function $G_J$ is no longer diagonal in the momentum space in the presence of a magnetic order and the exchange coupling.  We now treat the exchange coupling perturbatively for $(|JM| \ll |m|)$, where the system remains insulating at $\mu = 0$.  Then, we expand the Green's function $G_J$ using the unperturbed one $G_0(\bm{k},\omega) = [\omega - H_0(\bm{k})]^{-1}$.  When the magnetic order is written as 
\begin{equation}
\bm{M}(\bm{r}) = \sum_{\bm{Q}} \bm{M}(\bm{Q}) e^{i\bm{Q}\cdot\bm{r}}, 
\end{equation}
the equation for the polarization becomes 
\begin{align}
\Delta P_a &\simeq -\frac{e}{2} \sum_{\bm{Q}} \int \frac{d\omega}{2\pi} \int_{\bm{k}} \int_{0}^{J} d\tilde{J} \nonumber\\
&\quad\times \{ \tilde{J} \operatorname{tr}[ 
G_0(\bm{M}^*\cdot\bm{\sigma})G_0(\bm{Q}\cdot\bm{\alpha})G_0\alpha_a G_0(\bm{M}\cdot\bm{\sigma})G_0 
+ G_0(\bm{M}^*\cdot\bm{\sigma})G_0\alpha_a G_0(\bm{Q}\cdot\bm{\alpha})G_0(\bm{M}\cdot\bm{\sigma})G_0 \nonumber\\
&\hspace{34pt} + G_0\alpha_a G_0(\bm{M}^*\cdot\bm{\sigma})G_0(\bm{Q}\cdot\bm{\alpha})G_0(\bm{M}\cdot\bm{\sigma})G_0
- G_0\alpha_a G_0(\bm{M}\cdot\bm{\sigma})G_0(\bm{Q}\cdot\bm{\alpha})G_0(\bm{M}^*\cdot\bm{\sigma})G_0 \nonumber\\
&\hspace{34pt} - G_0(\bm{M}^*\cdot\bm{\sigma})G_0(\bm{Q}\cdot\bm{\alpha})G_0(\bm{M}\cdot\bm{\sigma})G_0\alpha_a G_0
+ G_0(\bm{M}\cdot\bm{\sigma})G_0(\bm{Q}\cdot\bm{\alpha})G_0(\bm{M}^*\cdot\bm{\sigma})G_0\alpha_a G_0 \nonumber\\
&\hspace{34pt} + G_0(\bm{M}\cdot\bm{\sigma})G_0(\bm{Q}\cdot\bm{\alpha})G_0\alpha_a G_0(\bm{M}^*\cdot\bm{\sigma})G_0
+ G_0(\bm{M}\cdot\bm{\sigma})G_0\alpha_a G_0(\bm{Q}\cdot\bm{\alpha})G_0(\bm{M}^*\cdot\bm{\sigma})G_0
] \},
\label{eq:S_polarization_perturb}
\end{align}
where $G_0 = G_0(\bm{k},\omega)$ and $\bm{M} = \bm{M}(\bm{Q})$.

\subsection{Model calculation}
\label{sec:S_polarization_1}

The polarization induced by the exchange coupling is obtained from Eq.~\eqref{eq:S_polarization_perturb} when the exchange coupling is treated perturbatively.  The present model contains the two coupling constants for the exchange coupling.  To evaluate the formula, we assume that the two coupling constants are proportional during the adiabatic insertion of the exchange coupling.  To be more specific, we put $J' = \chi J$, where $\chi$ remains constant while $J$ evolves.  Then, Eq.~\eqref{eq:S_polarization_perturb} leads to the uniform polarization per unit volume 
\begin{align}
\label{eq:S_polarization_isotropic}
\Delta\bm{P} &= -\frac{e\chi J}{6\pi^2 m} \sum_{\bm{Q}} \operatorname{Im} [\bm{M}_{\bm{Q}}^* (\bm{Q}\cdot\bm{M}_{\bm{Q}})] 
= -\frac{eJJ'}{6\pi^2 m} \sum_{\bm{Q}} \operatorname{Im} [\bm{M}_{\bm{Q}}^* (\bm{Q}\cdot\bm{M}_{\bm{Q}})]. 
\end{align}
When we use the real-space form $\bm{M}(\bm{r})$, this uniform polarization results in 
\begin{equation}
\Delta \bm{P} = \frac{eJJ'}{6\pi^2 m} \frac{1}{V}\int d\bm{r} \bm{M} (\nabla\cdot\bm{M}), 
\end{equation}
where $V$ is the volume of a unit cell and the spatial integration is performed over a unit cell determined by the magnetic order.  

We can rewrite the polarization in other forms.  Using the relation 
\begin{equation}
\bm{A}\times(\bm{B}\times\bm{C}) - (\bm{A}\times\bm{B})\times\bm{C} 
= \bm{A}(\bm{B}\cdot\bm{C}) - (\bm{A}\cdot\bm{B})\bm{C}
\end{equation}
or 
\begin{align}
\int d\bm{r} [ \bm{A} \times (\nabla \times \bm{B}) + \bm{B} \times (\nabla \times \bm{A}) ] 
= \int d\bm{r} [ \nabla (\bm{A} \cdot \bm{B}) -(\bm{A} \cdot \nabla) \bm{B} - (\bm{B} \cdot \nabla) \bm{A} ] 
= \int d\bm{r} [ \bm{A} (\nabla \cdot \bm{B}) + \bm{B} (\nabla \cdot \bm{A}) ], 
\end{align}
where we neglect the boundary contribution in the latter, the uniform polarization per unit volume becomes 
\begin{equation}
\Delta \bm{P} = -\frac{eJJ'}{6\pi^2 m} \sum_{\bm{Q}} \operatorname{Im} [\bm{M}_{\bm{Q}}^*\times(\bm{Q}\times\bm{M}_{\bm{Q}})], 
\end{equation}
or 
\begin{equation}
\Delta \bm{P} = \frac{eJJ'}{6\pi^2 m} \frac{1}{V}\int d\bm{r} \bm{M}\times(\nabla\times\bm{M}).   
\end{equation}
The real-space expressions $\bm{M}(\nabla\cdot\bm{M})$ and $\bm{M}\times(\nabla\times\bm{M})$ reminds us of the magnetoelectric effect, where magnetic structures that create a magnetic monopole or troidal moment yield finite effect.  For example, the former resembles the diagonal magnetoelectric effect, where the polarization is parallel to the magnetization with the coefficient proportional to the charge of the magnetic monopole.   
We note that we can see the similarity of the magnetic monopole and troidal moment because we now focus on the uniform polarization.

\begin{figure}
\centering
\includegraphics[width=0.5\hsize]{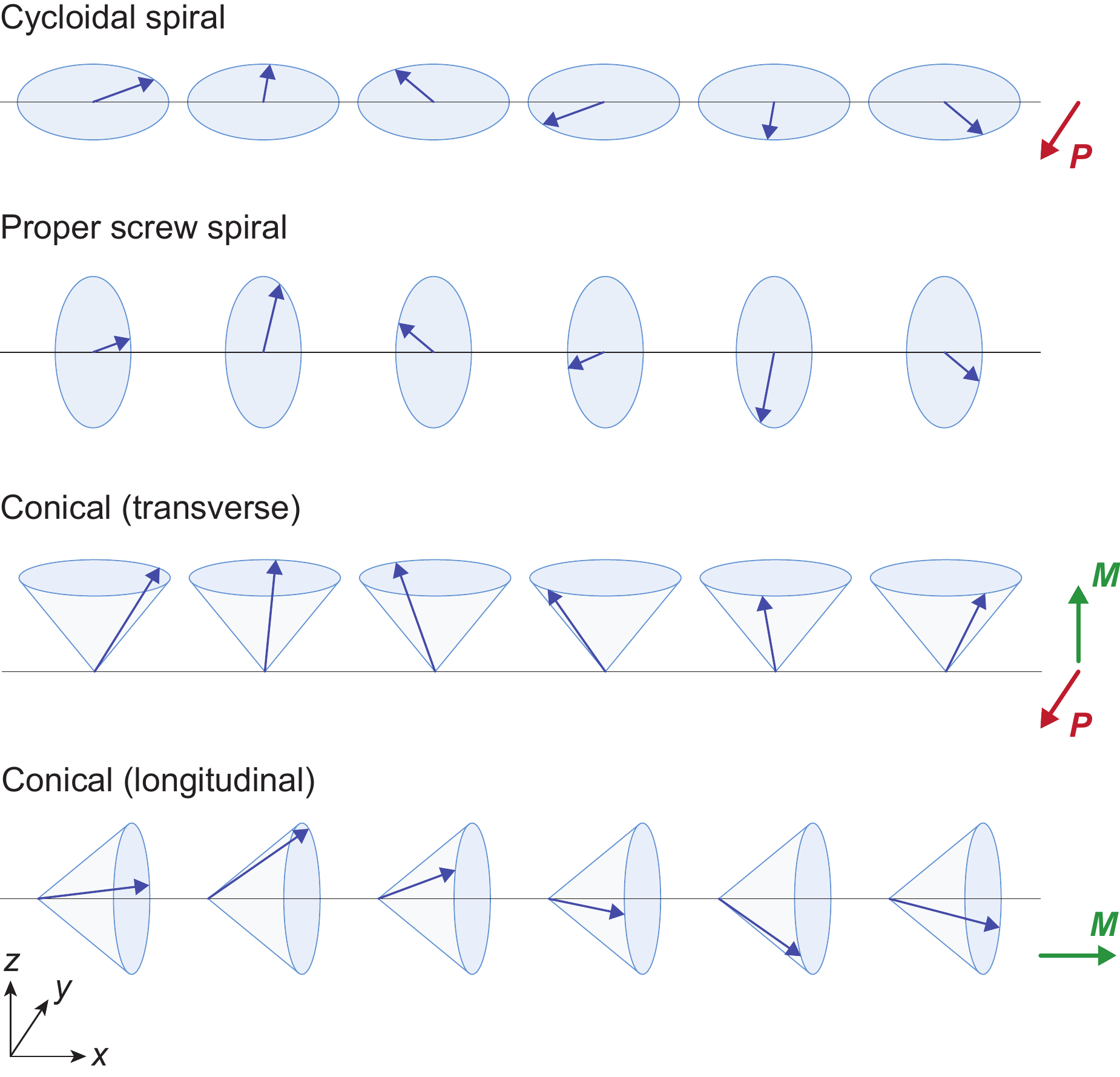}
\caption{
Spiral magnetic orders by local magnetic moments in an isotropic Dirac electron system.  The uniform polarization $\bm{P}$ induced by the exchange coupling between the magnetic moments and the Dirac electrons is depicted with the uniform magnetization $\bm{M}$. 
}
\label{fig:magnetic_order}
\end{figure}

Finite polarization requires that the magnetic order realize finite $\sum_{\bm{Q}} \operatorname{Im}[\bm{M}^*(\bm{Q}\cdot\bm{M})]$, which necessarily violates inversion.  This factor has the form $\int d\bm{r} [\bm{M}(\nabla\cdot\bm{M})] = -\int d\bm{r} [(\bm{M}\cdot\nabla)\bm{M}]$ in the real space and this is compatible with the Ginzburg--Landau theory obtained from the symmetry consideration \cite{S_Mostovoy}. 
In addition, the strength of the the exchange coupling must be different for the two constituent orbitals of the Dirac electrons, implied by the product $JJ'$.  
We suppose spiral spin orders with a single $\bm{Q}$ of the form
\begin{equation}
\bm{M}(\bm{r}) = M_1 \bm{e}_1 \cos (\bm{Q}\cdot\bm{r}) + M_2 \bm{e}_2 \sin (\bm{Q}\cdot\bm{r}) + M_3 \bm{e}_3, 
\end{equation}
where $\bm{e}_1$, $\bm{e}_2$, $\bm{e}_3$ are orthonormal, and illustrate typical cases in Fig.~\ref{fig:magnetic_order}.  The condition Eq.~\eqref{eq:S_polarization_isotropic} states that there is finite polarization when the plane spanned by $\bm{e}_1$ and $\bm{e}_2$ $(M_1, M_2 \neq 0)$ contains the wavevector $\bm{Q}$ and that the induced polarization is on the same plane but perpendicular to $\bm{Q}$.

The expression for the continuum model is to be contrasted with the atomic model \cite{S_Katsura}.  In the microscopic cluster model with two magnetic moments $\bm{S}_i$ and $\bm{S}_j$, the polarization becomes $\bm{P} \propto \bm{e}_{ij} \times (\bm{S}_i \times \bm{S}_j)$, where the vector $\bm{e}_{ij}$ connects the two magnetic moments.  Therefore, a noncolinear magnet hosts finite electric polarization.  
While our expression may be apparently different, we find the same result for uniform polarization for the spin textures listed in Fig.~\ref{fig:magnetic_order}.  
The microscopic model has an empty atomic site between the two magnetic moments and virtual transitions of an electron generate polarization.  
In our Dirac model, on the other hand, there are multiple bands from different elements that are extended in the bulk.  As the conduction and valence bands are related with the nontrivial band topology, a certain magnetic order gives rise to finite polarization.

\section{Estimates with material parameters}

\subsection{Review}

To evaluate the effects that we have obtained with realistic material parameters, we rewrite the results by recovering the Planck constant $h$ and the velocity $v$; the $\bm{k}\cdot\bm{p}$ Hamiltonian reads 
\begin{equation}
H_0(\bm{k}) = mv^2 \beta + \hbar v \bm{\alpha}\cdot\bm{k}.  
\end{equation}

In the insulating state, the electric polarization generated by an inversion-breaking magnetic order is 
\begin{equation}
\Delta\bm{P} = -\frac{eJJ'}{6\pi^2 \hbar v^3 |m|} \sum_{\bm{Q}} \operatorname{Im} [\bm{M}_{\bm{Q}}^* (\bm{Q}\cdot\bm{M}_{\bm{Q}})].  
\end{equation}

In the metallic state, we calculated the conductivity, a uniform magnetization, and the current induced by an oscillating magnetic order.  The conductivity is 
\begin{align}
\sigma_{ab}(\omega) 
&= \sigma_0(\omega) \delta_{ab} 
+ \sigma^\text{AH}_{ab}(\omega)
- \eta(\omega) \sum_{\bm{Q}} |\bm{M}_{\bm{Q}}|^2 \delta_{ab} 
+ \eta'(\omega) \sum_{\bm{Q}} \left( M^*_{\bm{Q},a} M_{\bm{Q},b} + M_{\bm{Q},a} M^*_{\bm{Q},b} \right),
\end{align}
with 
\begin{gather} 
\sigma_0(\omega) = \frac{e^2 |n(\epsilon_F)| \tau_\omega}{|m|}, \\
\eta(\omega) = \frac{2e^2 |n(\epsilon_F)| \tau_\omega^3}{|m| \hbar^2} (J+J'\tilde{\beta})^2, \\
\eta'(\omega) = \frac{e^2 |n(\epsilon_F)| \tau_\omega}{8|m|^3 v^4} (-3J^2 + 5J'^2 - 2JJ'\tilde{\beta}).  
\end{gather}
The magnetization is 
\begin{align}
\bm{m}_\text{Dirac} &= 
\lambda^{(1)} \sum_{\bm{Q}} (\bm{Q}\cdot\bm{E}) \operatorname{Im}(\bm{M}_{\bm{Q}}\times\bm{M}^*_{\bm{Q}}) 
+ \lambda^{(2)} \sum_{\bm{Q}} \Big\{\! \operatorname{Im}[(\bm{M}^*_{\bm{Q}}\times\bm{Q})(\bm{M}_{\bm{Q}} \cdot \bm{E})] 
+ \operatorname{Im}[(\bm{M}^*_{\bm{Q}}\times\bm{E})(\bm{Q}\cdot\bm{M}_{\bm{Q}})] \Big\}
\end{align}
with 
\begin{gather*}
\lambda^{(1)} = \frac{g\mu_B e n(\epsilon_F)}{\hbar m^2 v^2} \tau^3 J(J+J'\tilde{\beta}), \\
\lambda^{(2)} = \frac{g\mu_B e n(\epsilon_F)}{2\hbar m^2 v^2} \tau^3 (J^2-J'^2).
\end{gather*}
The current induced by an oscillating magnetic order is 
\begin{align}
\bm{j}(\omega) 
= &\sum_{\bm{Q}\omega_1\omega_2} \delta_{\omega_1+\omega_2,\omega} \Big[ \gamma^{(S)}(\omega_1,\omega_2) \bm{Q}\times(\bm{M}_{1}\times\bm{M}_{2}) \nonumber\\
&\quad + \gamma^{(A)}(\omega_1,\omega_2) 
\{ J [\bm{M}_{1}\times(\bm{Q}\times\bm{M}_{2}) + \bm{M}_{2}\times(\bm{Q}\times\bm{M}_{1})] 
+ J'\tilde{\beta} [\bm{M}_{1}(\bm{Q}\cdot\bm{M}_{2}) + \bm{M}_{2}(\bm{Q}\cdot\bm{M}_{1})] \} \Big],
\end{align}
with 
\begin{gather}
\begin{aligned}
\gamma^{(S)}(\omega_1,\omega_2) &= \frac{e}{8m^2 v^2} n(\epsilon_F) (J+J'\tilde{\beta})^2 \cdot
i(\omega_1+\omega_2)\tau_{\omega_1+\omega_2} ( \omega_1 \tau_{\omega_1}^2 + \omega_2 \tau_{\omega_2}^2 ),
\end{aligned}
\\
\gamma^{(A)}(\omega_1,\omega_2) = -\frac{e}{4m^2 v^2} (J+J'\tilde{\beta}) n(\epsilon_F) ( \omega_1 \tau_{\omega_1}^2 - \omega_2 \tau_{\omega_2}^2 ). 
\end{gather}

\subsection{Estimates for magnetically-doped TIs}

We adopt the values for the magnetically-doped TI Cr$_x$(Bi$_{1-y}$Sb$_y$)$_{2-x}$Te$_3$ in Ref.~\cite{S_Yasuda} presenting experiments of films, which uses the velocity $v = \SI{5.0e5}{m/s}$, the mass $m = -0.21 m_e$ converted from \SI{-300}{meV} with the electron mass $m_e = \SI{9.109}{kg}$, the coupling constants for the exchange coupling $J \mu_B = \SI{-5}{meV}$ and $J' \mu_B = \SI{1}{meV}$.  
In the following analysis for a metallic state, we set the Fermi energy $|\epsilon_F| = \SI{100}{meV}$, which is smaller than half the mass gap $|mv^2| = \SI{300}{meV}$. 
Then, Eq.~\eqref{eq:S_density} gives the carrier density $|n| \approx \SI{1.4e19}{cm^{-3}}$.  
We estimate the lifetime $\tau$ from the longitudinal conductivity $\sigma_0 \approx \SI{100}{\ohm^{-1}.cm^{-1}}$ using the relation $\sigma_0 = e^2 |n| \tau/|m|$ to obtain $\tau \approx \SI{5.4e-15}{s}$, which corresponds to an energy scale $\hbar/\tau \approx \SI{120}{meV}$.  
The RKKY interaction presumably plays a dominant role in forming a magnetic order in the metallic state with the wavenumber $Q$ being twice the Fermi wavenumber $2k_F$.  We use $Q = 2k_F \approx \SI{1.5e9}{m^{-1}}$ at $|\epsilon_F| = \SI{100}{meV}$.   
We suppose that each Cr atom has a magnetic moment $M = 3 \mu_B$ from the experimental observation.  

We evaluate our results using the material parameters above.  
In the metallic state, we first calculate the corrections to the conductivity.  The isotropic term similar to the magnetoresistance is 
\begin{align}
-\eta \sum_{\bm{Q}} |\bm{M}_{\bm{Q}}|^2 
\approx -\frac{2e^2 |n| \tau^3}{|m| \hbar^2} (J + J'\tilde{\beta})^2 \cdot 2M^2 
\approx 
\begin{cases}
\SI{-3.8}{\ohm^{-1}.\cm^{-1}} & (\tilde{\beta} = +1) \\
\SI{-8.7}{\ohm^{-1}.\cm^{-1}} & (\tilde{\beta} = -1)
\end{cases}, 
\end{align}
which reduces the conductivity about a few percent.   
The reduction depends on the bands labeled by $\tilde{\beta}$, reflecting the strength of the exchange coupling.  The magnitude of the $\eta'$ term, which resembles the anisotropic magnetoresistance and the planar Hall effect, is 
\begin{gather}
\eta' \cdot 4M^2 \approx \frac{e^2 |n(\epsilon_F)| \tau}{8|m|^3 v^4} (-3J^2 + 5J'^2 - 2JJ'\tilde{\beta}) \cdot 4M^2 \approx 
\begin{cases}
\SI{-0.3}{\ohm^{-1}.\cm^{-1}} & (\tilde{\beta} = +1) \\
\SI{-0.4}{\ohm^{-1}.\cm^{-1}} & (\tilde{\beta} = -1)
\end{cases}.
\end{gather}
It gives rise to the anisotropy in the conductivity.  

To estimate the magnetization induced by the electric field, we assume the current density $j = \SI{1e8}{A/m^2}$.  Then, the magnitude of the induced magnetization $m_\text{Dirac}$ is 
\begin{gather}
\lambda^{(1)} QM^2 \frac{j}{\sigma} \approx
\begin{cases}
\SI{4.5e-4}{A/m} \approx \SI{4.5e-7}{emu/cm^3} & (\tilde{\beta} = +1) \\
\SI{6.8e-4}{A/m} \approx \SI{6.8e-7}{emu/cm^3} & (\tilde{\beta} = -1)
\end{cases}, \\
\lambda^{(2)} QM^2 \frac{j}{\sigma} \approx \SI{2.7e-4}{A/m} \approx \SI{2.7e-7}{emu/cm^3}, 
\end{gather}
which may be tiny for an experimental observation.  We note that the induced magnetization is proportional to the current density.  

Lastly, we estimate the current density induced by an oscillating magnetic order.  When the oscillation frequency is \SI{1}{GHz}, namely $\omega = 2\pi \times 10^9\,\mathrm{rad/s}$, the sum frequency generation and the two terms for the difference frequency generation are 
\begin{gather}
\gamma^{(S)} \cdot QM^2 \approx 
\begin{cases}
\SI{4.1}{A/m^2} & (\tilde{\beta} = +1) \\
\SI{9.2}{A/m^2} & (\tilde{\beta} = -1)
\end{cases},\\
\gamma^{(A)} \cdot JQM^2 \approx 
\begin{cases}
\SI{-1.5e5}{A/m^2} & (\tilde{\beta} = +1) \\
\SI{-2.3e5}{A/m^2} & (\tilde{\beta} = -1)
\end{cases}, \\
\gamma^{(A)} \cdot J'\tilde{\beta}QM^2 \approx 
\begin{cases}
\SI{3.0e4}{A/m^2} & (\tilde{\beta} = +1) \\
\SI{-4.5e4}{A/m^2} & (\tilde{\beta} = -1)
\end{cases}, 
\end{gather}
where we set $\omega_1 = \omega_2 = \omega$ for the sum frequency generation and $\omega_1 = -\omega_2 = \omega$ for the difference frequency generation.  
With the oscillation frequency \SI{1}{GHz}, the sum frequency generation is much smaller than the difference frequency generation.  We note that the former grows quadratically with respect to the frequency while the latter does linearly.  At a higher frequency \SI{1}{THz} ($\omega = 2\pi \times 10^{12}\,\mathrm{rad/s}$), the response becomes 
\begin{gather}
\gamma^{(S)} \cdot QM^2 \approx 
\begin{cases}
\SI{4.1e6}{A/m^2} & (\tilde{\beta} = +1) \\
\SI{9.2e6}{A/m^2} & (\tilde{\beta} = -1)
\end{cases},\\
\gamma^{(A)} \cdot JQM^2 \approx 
\begin{cases}
\SI{-1.5e8}{A/m^2} & (\tilde{\beta} = +1) \\
\SI{-2.3e8}{A/m^2} & (\tilde{\beta} = -1)
\end{cases}, \\
\gamma^{(A)} \cdot J'\tilde{\beta}QM^2 \approx 
\begin{cases}
\SI{3.0e7}{A/m^2} & (\tilde{\beta} = +1) \\
\SI{-4.5e7}{A/m^2} & (\tilde{\beta} = -1)
\end{cases}.  
\end{gather}

In the insulating state, the wavenumber of a magnetic order may be different from that in the metallic state, but we use here the same value for the estimate.  Then, the electronic polarization induced by a spiral magnetic order becomes 
\begin{equation}
\Delta P \approx -\frac{eJJ'}{6\pi^2 \hbar v^3 |m|} QM^2 \approx \SI{1.8}{\micro\coulomb/\metre^2}. 
\end{equation}

In concluding the estimates, we comment on the surface contributions of magnetic topological materials.  In the metallic state, the surface contributions if present should be much smaller than the bulk contributions because of their small volume proportions.  In addition, the surface effect is insensitive to the sample thickness, which we could distinguish in experiments.  
On the other hand, when the bulk is in the insulating state, the surface of a magnetic topological material can be either metallic or insulating depending the Fermi energy.  Whether or not the surface contributes to the electric polarization, the electric polarization of the bulk depends on the sample volume or the thickness, whereas the surface contribution does not, which suggests an experimental identification.

\section{Stacking TI model}

\subsection{Model}

We turn to a stoichiometric magnetic TI, where magnetic elements are periodically aligned.  We consider the magnetic TI MnBi$_2$Te$_4$.  It consists of septuple layers stacking along the [0001] direction, bound by van der Waals forces.  Each septuple layer can be regarded as a TI and the surface states are coupled to the local magnetic moments of the periodic array of Mn.  
The effective model is founded on the topological surface states and it includes the coupling among the stacking layers and the magnetic moments \cite{S_MacDonald}: 
\begin{align}
\hat{H} = \sum_{\bm{k}_\perp,ll'} c_{\bm{k}_\perp l}^\dagger \{ [(-1)^l v \tau_z (\hat{z}\times\bm{\sigma})\cdot\bm{k}_\perp] \delta_{ll'} + \Delta_{ll'}(1-\delta_{ll'}) -\mu \} c_{\bm{k}_\perp l'} 
- J \sum_{\bm{r}} c_{\bm{r}}^\dagger [\bm{M}(\bm{r})\cdot\bm{\sigma}] c_{\bm{r}}. 
\end{align}
$\tau_{z} = \pm 1$ corresponds to the top or bottom surface state of each septuple layer, $\bm{\sigma}$ describes the spin degrees of freedom, $l$, $l'$ are layer indices, $\Delta_{ll'}$ is the strength of the interlayer hopping, and $J$ is the exchange coupling between the electrons and the local magnetic moments within a septuple layer.  We take the stacking direction as the $z$ axis and $\perp$ stands for the $xy$ plane.  
Electron hopping between layers are suppressed at a long distance: in the following, we include hopping between the top and bottom layers within a septuple layer $(\Delta_S)$ and between the nearest surface states of the adjacent layers $(\Delta_D)$.  

In the long-wavelength limit, the Hamiltonian except for the exchange coupling becomes
\begin{align}
H(\bm{k}) &= v \tau_z (\hat{z}\times\bm{\sigma})\cdot\bm{k}_\perp + \Delta_S \tau_x + \Delta_D (\tau_x \cos k_z d -\tau_y \sin k_z d) \nonumber\\
&\simeq (\Delta_S + \Delta_D) \tau_x + v \tau_z (\hat{z}\times\bm{\sigma})\cdot\bm{k}_\perp + (-\Delta_D d) k_z \tau_y . 
\end{align}
After rescaling the momentum, we henceforth use the Hamiltonian
\begin{gather}
H = m\beta + \bm{\alpha}\cdot\bm{\Pi}, \\
H' = -J\bm{M}(\bm{r},t)\cdot\bm{\sigma}, 
\end{gather}
where we define the matrices $\beta$ and $\bm{\alpha}$ by
\begin{equation}
\beta = \tau_x, \quad \bm{\alpha} = (-\sigma_y \tau_z, \sigma_x \tau_z, \tau_y).   
\end{equation}
They satisfy the anticommutation relations Eq.~\eqref{eq:S_anticommutation}.  
Here, we assume that every septuple layer has the same magnetic order.  A magnetic pattern along the stacking direction enlarges the matrix structure of the model.

\subsection{Polarization}
\label{sec:S_polarization_planar}

The crystalline structure of MnBi$_2$Te$_4$ belongs to the space group $R\bar{3}m$ (point group $D_{3d}$).  Since it possesses inversion symmetry, there is no electronic polarization without symmetry breaking.  However, when a spontaneous symmetry breaking occurs to develop a magnetic order that breaks inversion, the exchange coupling between the magnetic moments and electrons can induce finite electronic polarization.  

We calculate the uniform polarization induced by the exchange coupling using Eq.~\eqref{eq:S_polarization_perturb} to obtain 
\begin{gather}
\Delta \bm{P}_\perp = 
-\frac{eJ^2}{4\pi^2 m} \sum_{\bm{Q}} \operatorname{Im} (\bm{M}_{\bm{Q}}^* Q_z M_{\bm{Q},z}) 
= -\frac{eJ^2}{4\pi^2 m} \sum_{\bm{Q}} \operatorname{Im} [\bm{M}_{\bm{Q}}^*\times(Q_z\hat{z}\times\bm{M}_{\bm{Q}})], \\
\Delta P_z = 
\frac{eJ^2}{12\pi^2 m} \sum_{\bm{Q}} \operatorname{Im} [M_{z,\bm{Q}}^*(\bm{Q}_\perp\cdot\bm{M}_{\bm{Q}})] 
= \frac{eJ^2}{12\pi^2 m} \sum_{\bm{Q}} \operatorname{Im} [\bm{M}_{\bm{Q}}^*\times(\bm{Q}_\perp\times\bm{M}_{\bm{Q}})]_z. 
\end{gather}
The expressions have the similar structure as the isotropic case in Sec.~\ref{sec:S_polarization_1}, but they reflect the stacking structure of the material.

\subsection{Effective Hamiltonian}

By applying the formal expression Eq.~\eqref{eq:S_effective_formal}, we can obtain the effective Hamiltonian valid for large $m$:
\begin{align}
H_\text{eff} &= m\beta - e\Phi - J\bm{M}\cdot\bm{\sigma} + \frac{1}{2m}\beta (\bm{\Pi}\cdot\bm{\Pi} + e\bm{B}\cdot\bm{\Sigma}) \nonumber\\
&\quad + \frac{e}{8m^2} (\nabla\cdot\bm{E}) + \frac{e}{8m^2} [ \bm{\Pi}\cdot(\bm{\Sigma}\times\bm{E}) + (\bm{\Sigma}\times\bm{E})\cdot\bm{\Pi} ] \nonumber\\
&\quad + \frac{J}{8m^2} \Big[ (\bm{\Pi}\cdot\bm{\Pi} + e\bm{B}\cdot\bm{\Sigma}) (\bm{M}\cdot\bm{\sigma}) + (\bm{M}\cdot\bm{\sigma}) (\bm{\Pi}\cdot\bm{\Pi} + e\bm{B}\cdot\bm{\Sigma}) \nonumber\\
&\hspace{45pt} - 2 [(\hat{z}\times\bm{\sigma})\cdot\bm{\Pi}_\perp] (\bm{M}\cdot\bm{\sigma}) [(\hat{z}\times\bm{\sigma})\cdot\bm{\Pi}_\perp] - 2 \Pi_z (\bm{M}\cdot\bm{\sigma}) \Pi_z \nonumber\\
&\hspace{45pt}+ 2i\beta \left\{ [(\hat{z}\times\bm{\sigma})\cdot\bm{\Pi}_\perp] (\bm{M}\cdot\bm{\sigma}) \Pi_z - \Pi_z (\bm{M}\cdot\bm{\sigma}) [(\hat{z}\times\bm{\sigma})\cdot\bm{\Pi}_\perp] \right\}
\Big], 
\end{align}
where the matrix $\bm{\Sigma}$ is 
\begin{equation}
\bm{\Sigma} = (-\sigma_x \beta, -\sigma_y \beta, \sigma_z).  
\end{equation}
We also note $\bm{\alpha} = (\hat{z}\times\bm{\sigma})\tau_z + \hat{z}\tau_y$. 
The difference from the previous model Eq.~\eqref{eq:S_effectiveH} arise because of the different commutation relation between $\bm{\alpha}$ and $\bm{\sigma}$, which reflects the planar anisotropy of the present model.

\subsection{Current operator}

We can calculate the current response from the oscillation of the magnetic order by following the definition Eq.~\eqref{eq:S_current_definition}.  As the present model has a complication concerning the anisotropy, it would be useful to write down the general expression first.  The current operator $\bm{\mathcal{J}}$ obtained from the Hamiltonian \eqref{eq:S_effective_formal} is 
\begin{align}
\bm{\mathcal{J}} = -\frac{e}{m} \beta \bm{\Pi} - \frac{e^2}{4m^2} \bm{\Sigma}\times\bm{E} 
+ \frac{e}{4m^2} [ \bm{\Pi}H' + H'\bm{\Pi} - \bm{\alpha}H'(\bm{\alpha}\cdot\bm{\Pi}) - (\bm{\alpha}\cdot\bm{\Pi})H'\bm{\alpha} ] + O(m^{-3}).  
\end{align}
The first two terms correspond to the conventional current operator and the contribution from the spin-orbit coupling, respectively.  
By inserting the explicit form of $H'$, we obtain the current operator for the present model:  
\begin{align}
\bm{\mathcal{J}} &= -\frac{e}{m} \beta \bm{\Pi} - \frac{e^2}{4m^2} \bm{\Sigma}\times\bm{E} \nonumber\\
&\quad - \frac{eJ}{4m^2} \Big[ \bm{\Pi}(\bm{M}\cdot\bm{\sigma})
- (\hat{z}\times\bm{\sigma}) (\bm{M}\cdot\bm{\sigma}) [(\hat{z}\times\bm{\sigma})\cdot\bm{\Pi}_\perp] - \hat{z} \Pi_z (\bm{M}\cdot\bm{\sigma}) \nonumber\\
&\hspace{50pt} + i\beta \left\{ (\hat{z}\times\bm{\sigma})(\bm{M}\cdot\bm{\sigma})\Pi_z - \hat{z}(\bm{M}\cdot\bm{\sigma})[(\hat{z}\times\bm{\sigma})\cdot\bm{\Pi}_\perp] \right\} + \text{H.c.} \Big] + O(m^{-3}).  
\end{align}
Now we notice that the current operator couples to the magnetic order $\bm{M}$.  We can further calculate the Pauli matrices and the differential operators, but we conclude the calculation here; the focus of the section is to see that the electric current couples to the magnetic order.

\end{document}